\documentclass[preprint,12pt]{elsarticle}
\usepackage[T1]{fontenc}
\usepackage{amsmath,amssymb,amsfonts}        
\usepackage{mathptmx}
\usepackage[scaled=0.91]{helvet}
\usepackage{courier}
\usepackage{doi}
\usepackage{subfigure}
\usepackage{graphicx}
\usepackage{caption}
\usepackage{psfrag}
\usepackage{color}
\usepackage{xcolor}
\usepackage{hyperref}
\usepackage[numbers]{natbib}

\journal{Chaos, Solitons \& Fractals}

\begin{document}

\begin{frontmatter}
\title{Fractional Vegetation-Water Model \\
in Arid and Semi-Arid Environments: \\
Pattern Formation and Numerical Simulations
}

\author[]{Alessandra Jannelli}
\ead{ajannelli@unime.it}
\author[]{Maria Paola Speciale}
\ead{mpspeciale@unime.it}
\affiliation[]{ 
organization={Mathematical and Computer Sciences, Physical Sciences and Earth Sciences, \\ University of Messina},
addressline={Viale F. Stagno d'Alcontres 31}, 
postcode={98166}, 
city={Messina}, 
country={Italy}}

\begin{abstract}
In this paper, we present a { new space--}fractional mathematical model to describe the dynamics and the interaction between plants and water in arid and semiarid environments with and without slopes. {
By the Caputo fractional operator, the model allows to simulate the phenomena, related to the vegetation migration,  that occur in domains with different slopes.
 
The novelty of the study is to propose a new fractional model that, by assuming the fractional parameter linked to the slope of the domain, represents a connection between the Klausmeier model, where water advection occurs, to the Klausmeier-Gray-Scott model, where water diffuses. 
The term involving the space--fractional derivative describes an anomalous physical phenomenon that changes as the fractional parameter changes, modelling an anomalous water advection arising from the space nonlocality property of the fractional operator. An analytical study of the stability of the Hopf bifurcation demonstrates that the migration speed results to be a function of the fractional parameter, confirming the connection between the fractional parameter and the slope of the domain.
The oscillatory solutions and the vegetation pattern formation, obtained numerically, validate the analytical results and  confirm the reliability and efficiency of the fractional formulation of the considered model.}
\end{abstract}

\begin{highlights}
\item We propose a new fractional mathematical model, involving the Caputo derivative, that describes the dynamics and interaction between water and plants in arid and semiarid environments with and without slope;

\item The new proposed model represents a link between the Klausmeier and the Klausmeier Gray-Scott models;

\item The stability analysis and the Hopf bifurcation study of the migration speed are performed. 

\item The numerical solutions demonstrate the reliability of the proposed model, which guarantees the oscillatory solutions and the vegetation pattern formation.
\end{highlights}


\end{frontmatter}

\section{Introduction}
Vegetation patterns arise in arid and semi-arid regions \cite{Valentin_1999, Rietkerk_2004}, and it has been observed that, in a sloped domain, vegetation aligns in a banded pattern due to a downward-oriented flow of water \cite{Borgogno_2009}.
The formation of patterns has been observed as well in arid and semi-arid ecosystems with and without a slope \cite{Klausmeier_1999,Macfadyan_1950a,Macfadyan_1950b,Deblauwe_2008,Doelman_2013}.

The formation of such patterns is often modelled by reaction-diffusion equations, where an advection term accounts for the downward flow of water. One of the simplest and most commonly used models to study vegetation patterns on the sloped domain is the Klausmeier (KL) model \cite{Klausmeier_1999}, a system of two reaction-diffusion-advection partial differential equations that describes the evolution of water and plant biomass in arid and semi-arid environments in a two-dimensional domain.
It is the oldest and simplest of several continuous models for patterning due to water redistribution.

The dimensionless Klausmeier model (see \cite{Klausmeier_1999},\cite{Sherratt_2005}-\cite{Sherratt_2011}), in an one-dimensional domain, is the following
\begin{equation}\label{eqK}
\begin{cases}
\frac{\partial  u}{\partial t}  =     \frac{\partial^2  u}{\partial x^2} - m\, u+ \,u^2 w,\\
\frac{\partial  w}{\partial t}  = \nu \frac{\partial  w}{\partial x}     - w-u^2 w+a.
\end{cases}
\end{equation}

It describes the interaction between plants $u(t,x)$ and water $w(t,x)$ in arid and semi-arid environments. The parameters $a$, $m$ and $\nu$ are constant and
linked to rainfall, plant loss, due to the death rate of vegetation, and slope gradient in the dimensionless quantities, respectively.
The linear term $-m\,u$ represents the mortality of plant biomass and the nonlinear term $u^2 w$ represents the water uptake by plants. This nonlinear term also appears in the second equation and describes the plant growth, i.e., water uptake by vegetation is converted into plant biomass at a constant rate. The one-dimensional domain has a constant slope, is generally gentle, and is oriented so that \lq \lq uphill\rq \rq \ corresponds to the direction of increasing $x$. The parameters $a$, $m$ and $\nu$, characterizing different ecosystems, are assumed such that $\nu \gg  1$, slope gradient, is taken such that the downhill advection term is large and $a$ and $m$ are taken in the ranges $[0.1,3.0]$ and $[0.05,2.0]$, respectively \cite{Klausmeier_1999,Rietkerk_2002}.

Now, we consider a reformulated version of the Klausmeier model, where the advection term of the water is replaced with a diffusion term, describing the plants' growth on flat land instead of hills. The dimensionless model is written as follows
\begin{equation}\label{eqGS}
\begin{cases}
\frac{\partial  u}{\partial t}  =     \frac{\partial^2  u}{\partial x^2} - m\, u+ \,u^2 w,\\
\frac{\partial  w}{\partial t}  = \nu \frac{\partial^2  w}{\partial x^2}   - w-u^2 w+a.
\end{cases}
\end{equation}
The model is a system of two reaction-diffusion equations, known as Gray--Scott one, generally applied in chemistry to describe the interaction between two concentrations of reacting substances,
substrate and activator. It is also used to describe dynamical processes of a non-chemical nature. Examples are found in biology, geology, physics, ecology and so on 
\cite{Murray_2002}-\cite{Meinhardt_1982}.
In this context, the system (\ref{eqGS}), also known as the diffusion Klausmeier--Gray--Scott model (KL--GS), describes the interaction between the water concentration $w$ (substrate) and the plant density $u$ (activator).
Under this interpretation, precipitation increases water concentration uniformly across space at a constant rate $a$ and water is converted to a plant density at a rate $- u^2 \,w$ or lost by evaporation at a rate normalized to $-w$. In addition to being generated via water, the plant dies at a rate $-m\, u$. Finally, both the plant and water spread through the space with the water diffusing $\nu$ times faster.
Here, the parameter $\nu \gg  1$, due not to the slope, is large with respect to the diffusion plant coefficient.



In the Klausmeier model, the parameter $\nu$ is large because it reflects the relative rates of water flow downhill and plant dispersal. In the model (\ref{eqGS}), since the plants grow on flat land instead of the hill, the diffusion effect arises instead of the advection one.
The behavior of the solutions of the models (\ref{eqK}) and  (\ref{eqGS}) has been studied in a lot of papers \cite{Sherratt_2007,Doelman_1997,Morgan_2000},\cite{Consolo_2017}-\cite{Consolo_2024}, in which pattern solutions are investigated in various parameter regimes.

{
The vegetation pattern formation in arid and semi-arid ecosystems can change, influenced by soil, climate, temperature and other environmental factors like a wide variety of locations upstream (i.e., space nonlocality).  The spatial non-locality is due to the high variation and long-range dependence of the slope of the terrains.
Fractional derivatives are more suitable for describing real-life applications than integer derivatives because of their non-locality. The locality property of integer derivatives can determine some limitations in describing such a formation process.
}
In the literature, fractional derivatives have been widely used to describe some physical and chemical phenomena with anomalous diffusion and/or advection processes and the boundary layer flow of viscous fluid \cite{Henry_2000}-\cite{Chen_2015}. The mathematical models that describe these problems are usually fractional advection-reaction-diffusion models \cite{JS_AIP_2019}-\cite{Gafiychuk_2006} such as the fractional version of the well-known Gierer-Meinhardt \cite{JS_2023}, Biswas–Milovic \cite{Ahmadian_2016} and also with the well-known Blasius model \cite{JS_2024}. 
Furthermore, the fractional diffusion equations have been shown to be quite efficient in describing the diffusion in complex systems by operators whose main particularity is their non-local behaviour.

In this paper, we propose a { new} fractional mathematical model describing the dynamics and interaction of plants and water on flat and no-flat domains. In particular, by using the { space} fractional Caputo operator, we develop a model to describe the anomalous physical processes that ensure the vegetation patterns formation. 
The main aim of this article is to demonstrate how the new fractional model allows us to describe the behavior of the solution 
by varying only the value of the fractional parameter, { linked to slope of the domains, preserving the vegetation pattern formation, whose migration speed is found by an analytical study of the stability of the Hopf bifurcation. Numerical results complete the presented study.}

In Section $2$, we recall the travelling wave solution to reduce the KL and KL--GS models to systems of ordinary differential equations, reporting the stability analysis concerning the Hopf bifurcation. In Section $3$, we present the new fractional model. We perform the stability analysis of the equilibrium points and we find the values of the migration speed, at which the Hopf bifurcation occurs and the pattern formation is guaranteed. { We find that the migration speed is a function of the parameter $\alpha$ and varies with the slope.} In Section $4$, we present the numerical method to solve the new fractional model, proving the pattern formation for various values of the fractional parameter. 
The obtained numerical solutions validate the analytical results and confirm the reliability and effectiveness of the proposed model. { The last Section is devoted to conclusions and future works.}

\section{Traveling wave solution and reduced models}
In this Section, we reduce both models (\ref{eqK}) and (\ref{eqGS}) into systems of ordinary differential equations by introducing the traveling wave solution. We recall the procedure proposed by Sherratt et al. \cite{Sherratt_2011} to study the stability 
of the KL model and to hold that patterns occur for a range of values of parameters ($a$, $m$ and $\nu$), bounded by a Hopf bifurcation point. 

The KL model (\ref{eqK}) admits the travelling wave solution
\begin{eqnarray}\label{tras}
u(t,x)=U(x-c t)=U(z) \qquad   w(t,x)=W(x-c t)=W(z),
\end{eqnarray}
where $c>0$ is the migration speed in the uphill direction.
In terms of new variables, the model is rewritten into the following system of ordinary differential equations
\begin{eqnarray}\label{eqK0}
\begin{cases}
U^{\prime \prime}+c U^{\prime}-m U+ U^2 W=0  \\
(\nu+c) W^{\prime}-W- U^2 W+a=0  .
\end{cases}
\end{eqnarray}
When $a > 2m$ with $m<2$, the model admits three uniform steady states whose behavior depends upon the values of the parameters. The previous conditions of the parameters $a$ and $m$ ensure that the rainfall is large enough to sustain plant growth in arid and semi-arid environments.
The equilibrium points are
\begin{equation}
(U_0,W_0)=(0,a), \qquad (U_{\pm}, W_{\pm}) =\left( \frac{a \pm \sqrt{a^2 - 4 m^2}}{2m},\frac{a \mp \sqrt{ a^2 - 4 m^2}}{2}\right). \label{equ_points}
\end{equation}
$(U_0, W_0)$ is a stable point and leads to the desertification.  $(U_{-}, W_{-})$  is a saddle point, whereas the point $ (U_{+}, W_{+})$ depends on the involved parameters \cite{Sherratt_2011}. The Turing-type patterns and, consequently, the existence of periodic solutions arise for homogeneous perturbations of the point $ (U_{+}, W_{+})$.

Concerning the reduced model (\ref{eqK0}),
in \cite{Sherratt_2007}-\cite{Sherratt_2011}, it proves that patterns occur for a suitable range of values of the parameters $a$ and $m$,  for large values of the slope parameter $\nu$ with a migration speed $c$ close to the maximum admitted value. 
Thought a rescaling of coordinate, $\zeta = z/{\nu}$, $U=\bar{U}(\zeta)$ and $W=\bar{W}(\zeta)$, we get
\begin{eqnarray}
\begin{cases}
\displaystyle{\frac{\bar{U}^{\prime\prime}}{\nu^2}}+\frac{l}{\nu}\bar{U}^{\prime}- m\bar{U}+ \bar{U}^2 \bar{W}=0 \\
\\
\left (1+ \displaystyle{\frac{c}{\nu}}\right)\bar{W}^{\prime}-\bar{W}- \bar{U}^2 \bar{W}+a=0  \end{cases}
\end{eqnarray}
where $\frac{1}{\nu^2}$, for large values of $\nu$, is very small and the order of magnitude of $c$ is assumed the same as of $\nu$. By applying the following transformation
\begin{equation}
\tilde{U}=\bar{U}, \quad \tilde{W}=\left(1+\frac{\nu}{c}\right)\bar{W}, \quad \tilde{\zeta}=\left(1+\frac{c}{\nu}\right)^{-1}\zeta, 
\end{equation}
the system (\ref{eqK0}) is mapped into
\begin{eqnarray}
\begin{cases}
\tilde{U}^{\prime}-M \tilde{U}+ \tilde{U}^2 \tilde{W}=0  \\
\tilde{W}^{\prime}-\tilde{W}- \tilde{U}^2 \tilde{W}+A=0  \label{eqSher},
\end{cases}
\end{eqnarray}
being $M=(1+\frac{\nu}{c})m$ and $A=(1+\frac{\nu}{c})a$. The second derivative in the first equation has been omitted since for large values of $\nu$,  $\displaystyle{\frac{1}{\nu^2}}$ it is very small so that the diffusion phenomenon of plants is negligible.

The equilibrium states of the system (\ref{eqSher}) are still $(\tilde{U}_0,\tilde{W}_0)=(0,A)=(U_0,W_0)$ and
\begin{equation}
 (\tilde{U}_{\pm}, \tilde{W}_{\pm}) =\left(\frac{A \pm \sqrt{A^2 - 4 M^2}}{2M},\frac{A \mp \sqrt{ A^2 - 4 M^2}}{2}\right)=
 (U_{\pm},W_{\pm}).
\end{equation}
Starting from the characteristic polynomial  
\begin{equation}\lambda^2+(M-1-U_{\pm}^2)\lambda+M(U_{\pm}^2-1)=0,\label{condpol0}
\end{equation}
 an analysis of stability for $(U_{\pm}, W_{\pm})$ has been developed in \cite{Sherratt_2011}, in particular,  for the state $(U_{+},W_{+})$ has been obtained the Hopf bifurcation point, initiating a branch of pattern
(periodic) solutions.

We recall that an equilibrium point is unstable if and only if complex and conjugate eigenvalues ($\lambda_1$, $\lambda_2$) exist such that
\begin{equation}\label{argl}
|\arg(\lambda_{1,2})|=\left|\arctan{\left(\frac{\sqrt{4M(U_{+}^2-1)-(M-1-U_{+}^2)^2}}{-M+1+U_{+}^2}\right)}\right|\le \frac{\pi}{2},
\end{equation}
with \begin{equation}
M-1-U_{\pm}^2\leq 0 \qquad 
\Delta=(M-1-U_{+}^2)^2-4M(U_{+}^2-1)<0. \label{cond00}
\end{equation}
The Hopf bifurcation is given by the equality of the above condition (\ref{argl}) that can be verified only by requiring that 
\begin{equation}
{U_{+}}^2 =M- 1 \qquad \Delta=4M(2-M)<0\quad  \rightarrow \quad M>2, \label{cond}\end{equation}
so that the eigenvalues are purely imaginary.  For suitable choices of the parameters such that the conditions (\ref{cond}) are verified, the migration speed value,  $c = c^{HB} $, for which the Hopf bifurcation occurs, is given by
\begin{equation} 
c=c^{HB}=\frac{\nu m}{U_{+}^2+1-m} \quad \mbox{and} \quad c<\frac{\nu m}{2-m}, \label{condHB}
\end{equation}
obtained by solving (\ref{cond})$_1$ and (\ref{cond})$_3$  with respect to $c$ in terms of the original model parameters. 
This means that, fixed $a$ and $m$, with $a > 2m$  and $m<2$, the maximum speed $c^{HB}$ is proportional to the slope parameter $\nu$ and increases with plant loss parameter $m$. The branch of pattern solutions terminates at a homoclinic solution, moving at a constant speed $c$.
For homogeneous perturbations of $ (U_{+}, W_{+})$, the origin of patterns is bounded by a Hopf bifurcation point given in (\ref{condHB}) 
that can be subcritical (or supercritical) according to $ M=(1+\frac{\nu}{c})m>4$ ( or $ M=(1+\frac{\nu}{c})m<4$) (see \cite{Sherratt_2011}).
{ Moreover, in \cite{Sherratt_2011}, it was demonstrated that for any $M >2$, there is a value of $A>2M$  at which the equations (\ref{eqSher}) have a solution that is homoclinic to the steady state $(U_{-}, W_{-})$.}

As far as the KL--GS model (\ref{eqGS}), the traveling wave solution (\ref{tras}) is admitted, then the model reduces to the following one
\begin{eqnarray}\label{eqGS0}
\begin{cases}
U^{\prime \prime}+c U^{\prime}-m U+ U^2 W=0    \\
\nu W^{\prime \prime}+c W^{\prime}-W- U^2 W+a=0   .
\end{cases}
\end{eqnarray}
It admits equilibrium points (\ref{equ_points}) which have, as far as stability is concerned, a similar behavior as the states of the KL model. In this context, we are interested in the formation of stable patterns with spatially periodic solutions, which occurs when the migration velocity $c$ is equal to zero \cite{Doelman_2013,Doelman_1997,Doelman_1998,Gandhi_2018} and for suitable values of parameters $a$, $m$ and $\nu$. Starting from the characteristic polynomial 
\begin{equation}
\nu \lambda^4-\lambda^2(U_+^2+1-m \nu)+ m(U_+^2-1)=0   
\end{equation}
relative to the first order system associated to the (\ref{eqGS0}) for the
state (\ref{equ_points}), the Hopf bifurcation point, initiating a branch of
 spatially pattern (periodic) solution, is obtained when the eigenvalues are purely imaginary (see Sect. 2 \cite{Gandhi_2018}).  We recall that the model (\ref{eqGS0}) describes the dynamics in a flat environment, and therefore the water diffusion process, involved in the KL--GS model, is preserved.

In the next Section, starting from the study of the above models (\ref{eqK0}) and (\ref{eqGS0}), we introduce a new fractional model that links the reduced KL model to the reduced KL--GS one with the aim to obtain  oscillatory solutions that preserve the vegetation pattern formation as the parameter $\alpha$ varies.

\section{The New Proposed Fractional Model}

{
The vegetation pattern formation in arid and semi-arid ecosystems is influenced by soil, climate, temperature, and other environmental factors. 
One of the most relevant environmental factors is the 
 wide variety of locations upstream (i.e., space non-locality) due to the high variation and long-range dependence of the slope of the terrains.
The locality property of integer derivatives can determine some limitations in describing such a formation process.
Instead, due to their non-locality, the fractional derivatives are more suitable for describing real-life applications than integer derivatives.
We propose a new mathematical problem by introducing the fractional derivative and show how the migration of vegetation patterns is influenced by the slope of the domain. 

It is well known that vegetation patterning migrates upward along the slope direction due to the downhill flow of the water. Some studies have shown that the rate of upward migration is related to the slope size \cite{Rietkerk_2002}, \cite{Consolo_2017}-\cite{Consolo_2022}. 
When the domain is flat, there is no vegetation pattern migration due to the absence of the transport of water; otherwise, when the domain is sloped, the vegetation migration process occurs due to the downhill water flow.
These phenomena are described by KL--GS and KL models, respectively.
In this study, we demonstrate that the migration of vegetation patterns is linked to the slope rate of the domain by the fractional order of the derivative. The use of the fractional derìvative in the new model leads to an anomalous physical phenomenon, as confirmed by the numerical results in Section  $4$.
}

{
Among the wide class of fractional operators known in the literature, we propose the Caputo derivative. The main advantage of this operator is that the initial conditions for fractional differential equations take the same form as the one for integer-order differential equations, i.e. contain the limit values of integer-order derivatives of unknown functions at the lower limit on the integration domain. Moreover, the numerical method, that we use in the following, is a good approach for this kind of operator. }

\bigskip

Taking into account the properties of the fractional order operators and the physical meaning of the dynamical processes described by (\ref{eqK}) and  (\ref{eqGS}), starting from  (\ref{eqK0})
we propose the following fractional model (FM)
\begin{eqnarray}\label{eqF}
\begin{cases}
 U^{\prime \prime}+c U^{\prime}-m U+ U^2 W=0   \\
 \nu D_z^{\alpha+1}W+c W^{\prime}-W- U^2 W+a=0  ,
\end{cases}
\end{eqnarray}
obtained by replacing the first derivative of the water density with respect to $z$ variable, appearing in the second equation of the models (\ref{eqK0}) 
by the fractional derivative in terms of the Caputo operator, $D_z^{\alpha+1}$, defined as follows \cite{Podlubny99, Diethelm_2004} 
\begin{equation*}
D_z^{\alpha+1} W(z)=
\begin{cases}
D_z^\alpha\left( W'(z) \right) =
\frac{1}{\Gamma (1-\alpha)}\int^z_0 W''(s)  (z-s)^{-\alpha}d s \quad 0\le \alpha < 1\\
W''(z) \qquad \qquad \qquad \qquad \qquad \qquad \qquad  \qquad \quad \alpha=1,
\end{cases}
\end{equation*}
where $\Gamma(\cdot)$ is the Euler's gamma 
function and $\alpha$ represents the fractional derivative order. It is worth noting that, for $\alpha=0$ the fractional model (\ref{eqF}) reduces to (\ref{eqK0}), otherwise for $\alpha=1$ it reduces to (\ref{eqGS0}).

\bigskip

The key point of the proposed formulation, model (\ref{eqF}), stands in to assume that the parameter $\alpha$ of the fractional operator is linked to the slope of the domains, { so that the new fractional model can describe the migration in the uphill direction related to domains with any slope.}
{ We note that this assumption is validated by the analytical study performed on the Hopf bifurcation of the migration speed.
In fact, in the next Section, we show that the migration speed $c$, at which the solutions move, depends on the fractional parameter $\alpha$. So that,
as the slope approaches zero, the migration speed $c$ approaches zero, according to the real phenomena.

Note that for the definition of the Caputo derivative, for $\alpha=0$, the fractional derivative is the first-order derivative and we recover the reduced KL model (\ref{eqK0}) that describes a classical advection process. For $\alpha=1$ the fractional derivative is
the second-order derivative and we recover the reduced KL--GS model (\ref{eqGS0}) describing the classical diffusion process.
We are interested to study the proposed model for $0<\alpha<1 $, taking into account that, when $\alpha  \rightarrow 0$ the water moves downhill (typical process in sloped domains) and when $\alpha \rightarrow 1$, the water tends to diffuse (typical process in flat domains). 
Finally, in this way, we demonstrate that
the new fractional model represents a connection from the KL model to the KL--GS one, as the parameter $\alpha$ varies.}
The fractional formulation of the mathematical model allows to obtain the  oscillatory behavior of the solutions and, therefore, ensures the vegetation patterns formation in arid and semi-arid environments with any slope.

\bigskip

In the following, we perform the stability analysis of the proposed model (\ref{eqF}) only for the unstable equilibrium point for which the vegetation pattern formation occurs. 

\bigskip

We apply the same procedure to study the stability of equilibrium points of a proposed model (\ref{eqF}), the approach applied to the system (\ref{eqK0}) in the previous section. By the rescaling of coordinate, $\zeta =z/{\nu}$, $U=\bar{U}(\zeta)$ and $W=\bar{W}(\zeta)$, the fractional model (\ref{eqF}) is mapped into the following one
\begin{eqnarray*}
\begin{cases}
\frac{\bar{U}^{\prime\prime}}{\nu^2}+\frac{c}{\nu}\bar{U}^{\prime}- m\bar{U}+ \bar{U}^2 \bar{W}=0 \\
\frac{1}{\nu^\alpha} D_{\zeta}^{\alpha+1}\bar{W}+\frac{c}{\nu}\bar{W}^{\prime}-\bar{W}- \bar{U}^2 \bar{W}+a=0  .
\end{cases}
\end{eqnarray*}
By adding and subtracting the term $(1-\alpha)\bar{W}^{\prime} $ in the second equation, we obtain
\begin{eqnarray}\label{eqF01}
\begin{cases}
\frac{\bar{U}^{\prime\prime}}{\nu^2}+\frac{c}{\nu}\bar{U}^{\prime}- m\bar{U}+ \bar{U}^2 \bar{W}=0 \\
\frac{1}{\nu^\alpha} D_{\zeta}^{\alpha+1}\bar{W}-(1-\alpha)\bar{W}^{\prime} +(1-\alpha+\frac{c}{\nu})\bar{W}^{\prime}-\bar{W}- \bar{U}^2 \bar{W}+a=0 , 
\end{cases}
\end{eqnarray}
where
$$\frac{1}{\nu^\alpha} D_{\zeta}^{\alpha+1}\bar{W}-(1-\alpha)\bar{W}^{\prime}=\left\{\begin{array}{ll} 0 & \alpha= 0 \\ \frac{1}{\nu} D_{\zeta}^{\prime \prime}\bar{W}  & \alpha = 1 \end{array}\right. $$ 
with 
$$
\left | \frac{1}{\nu^\alpha} D_{\zeta}^{\alpha+1}\bar{W}-(1-\alpha)\bar{W}^{\prime}\right | \ll 1
$$
for large values of $\nu$ and different values of $\alpha$, with $0<\alpha<1$. Now, we set
\begin{equation}
\left | \frac{1}{\nu^\alpha} D_{\zeta}^{\alpha+1}\bar{W}-(1-\alpha)\bar{W}^{\prime} \right |= k_\alpha \label{cond_k}
\end{equation}
assuming that $k_\alpha$  is very small, positive and depending on $\alpha$ so that we are able to study the stability and the Hopf bifurcation of the proposed system as a first-order one. 
The assumption on the values of $k_\alpha$ will be confirmed, in the next Sections, by the numerical results.
Moreover, the term involving $\frac{1}{\nu^2}$, in the first equation, is negligible due to large values of $\nu$ and  
the order of magnitude of $c$ is the same as of $\nu$ (see Sherratt \cite{Sherratt_2011}).

By applying the following transformation to the system (\ref{eqF01})
\begin{equation}
\tilde{\tilde{U}}=\bar{U}, \quad \tilde{\tilde{W}}=\left(1-\alpha+\frac{\nu}{c}\right)\bar{W}, \quad \tilde{\tilde{\zeta}}=\left(1-\alpha+\frac{c}{\nu}\right)^{-1}\zeta,
\end{equation}
we get 
\begin{eqnarray}\label{sys1}
\begin{cases}
 \tilde{\tilde{U}}^{\prime}- M_{\alpha}  \tilde{\tilde{U}}+  \tilde{\tilde{U}}^2  \tilde{\tilde{W}}=0  \\
 \tilde{\tilde{W}}^{\prime}- \tilde{\tilde{W}}-  \tilde{\tilde{U}}^2  \tilde{\tilde{W}}+ A_\alpha+K_\alpha=0,  \end{cases}
\end{eqnarray}
with 
\begin{equation*}
M_{\alpha}=\left(1+\frac{(1-\alpha)\nu}{c}\right)m, \qquad  A_{\alpha}=\left(1+\frac{(1-\alpha)\nu}{c}\right)a, \qquad K_{\alpha}=\left(1+\frac{(1-\alpha)\nu}{c}\right)k_\alpha.
\end{equation*}
The equilibrium states of the system (\ref{sys1}) are $( \tilde{\tilde{U}}_0, \tilde{\tilde{W}}_0)=(0,A_\alpha)$ and
\begin{equation*}
 ( \tilde{\tilde{U}}_{\pm},  \tilde{\tilde{W}}_{\pm}) =\left(\frac{A_\alpha+K_\alpha\pm \sqrt{ (A_\alpha+K_\alpha)^2 - 4  M_\alpha^2}}{2 M_{\alpha}},\frac{ A_\alpha+K_\alpha \mp \sqrt{  (A_\alpha+K_\alpha)^2 - 4  M_{\alpha}^2}}{2}\right).
\end{equation*}
For $\alpha=0$ and $\alpha=1$, it holds
\begin{equation*}
( \tilde{\tilde{U}}_0, \tilde{\tilde{W}}_0)=(U_0,W_0) \qquad \qquad ( \tilde{\tilde{U}}_{\pm},  \tilde{\tilde{W}}_{\pm}) =(U_{\pm},W_{\pm}).
\end{equation*}
The characteristic polynomial is given by
\begin{equation}\lambda^2+( M_{\alpha}-1- \tilde{\tilde{U}}_{\pm}^2)\lambda+ M_{\alpha}( \tilde{\tilde{U}}_{\pm}^2-1)=0\label{condpol}.
\end{equation}
In this context, we are interested to find the Hopf bifurcation point for the unstable state $(\tilde{\tilde{U}}_{+}, \tilde{\tilde{W}}_{+})$.
In this case,  an equilibrium point is unstable if and only if eigenvalues ($\lambda_1$, $\lambda_2$) exist such that
$$|\arg(\lambda_{1,2})|=\left|\arctan{\left(\frac{\sqrt{4M_{\alpha}(\tilde{\tilde{U}}_{+}^2-1)-(M_{\alpha}-1-\tilde{\tilde{U}}_{+}^2)^2}}{-M_{\alpha}+1+\tilde{\tilde{U}}_{+}^2}\right)}\right|\le\frac{\pi(1-\alpha)}{2}\leq \frac{\pi}{2},$$
(for $\alpha=0$ we get (\ref{argl})), and when 
\begin{equation}
{M_{\alpha}-1- \tilde{\tilde{U}}_{+}^2}\leq 0, \quad 
\Delta=(M_{\alpha}-1- \tilde{\tilde{U}}_{+}^2)^2-4M_{\alpha}( \tilde{\tilde{U}}_{+}^2-1)<0\label{delta}.
\end{equation}
The above relations are verified for the (\ref{cond00}), (\ref{cond}) and for the following conditions
\begin{eqnarray*}
&& M_{\alpha}-1-\tilde{\tilde{U}}_{+}^2\le M-1-U_{+}^2\le 0\\
&&(M_{\alpha}-1-\tilde{\tilde{U}}_{+}^2)^2-4M_{\alpha}(\tilde{\tilde{U}}_{+}^2-1)\le(M-1-U_{+}^2)^2-4 M(U_{+}^2-1)<4 M(2-M)<0.
\end{eqnarray*}
We obtain the Hopf bifurcation by requiring the following equality condition 
\begin{equation*}
|\arg(\lambda_{1,2})|=\left|\arctan{\left(\frac{\sqrt{4M_{\alpha}(\tilde{\tilde{U}}_{+}^2-1)-(M_{\alpha}-1-\tilde{\tilde{U}}_{+}^2)^2}}{-M_{\alpha}+1+\tilde{\tilde{U}}_{+}^2}\right)}\right|=\frac{\pi(1-\alpha)}{2}
\end{equation*}
by which we get the migration speed, $c=c_\alpha^{HB}$ for the state $(\tilde{\tilde{U}}_{+}, \tilde{\tilde{W}}_{+})$
\begin{eqnarray}  \label{condHBF}
&&c_\alpha^{HB}=   \frac{(1-\alpha) m \nu}{(1-m+\tilde{\tilde{U}}^2)^2+2 m \left(\tilde{\tilde{U}}_{+}^2-1\right) (\cos (\pi  \alpha)-1)} \nonumber \\
&& \qquad\left(1-m+\tilde{\tilde{U}}_{+}^2+(\cos (\pi  \alpha)-1)(1-\tilde{\tilde{U}}_{+}^2)+ \right.\\ 
&&\qquad\left.-\sqrt{\left(\tilde{\tilde{U}}_{+}^2-1\right) \left(2 \tilde{\tilde{U}}_{+}^2(1-\cos(\pi \alpha))^2+(\tilde{\tilde{U}}_{+}^2+1) \sin^2\left(\pi  \alpha\right)\right)}\right). \nonumber
 \end{eqnarray}
We note that when $\alpha= 0$, we find again the condition (\ref{condHB}) of the migration speed and the conditions for the parameters for the fractional KL model, and, when $\alpha \rightarrow 1$, we get the migration speed such that $c \rightarrow 0$.
From the critical value of migration speed, $c_\alpha^{HB}$,  emerges a branch of the patterns that will be confirmed in the next section by the numerical results.
In the following, we set the parameters such that stable patterns arise bound by subcritical Hopf bifurcation, that occurs when  $ \left(1-\alpha+\frac{\nu}{c}\right)m>4$.

\subsection{Examples of physical applications}

With the aim to preserve the formation and dynamics of the vegetation patterns, in the following, we choose to set, as an example, the values of parameters $a=2$ and $m=0.45$ and only change the value of $\nu$, so that $(U_{+}, W_+)$ is unstable and the pattern formation of the models KL and KL--GS is guaranteed. We have that the formula (\ref{condHB}) and the formulas (2.6) in Sect. 2 of \cite{ Gandhi_2018} are satisfied. 
The coordinates of the equilibrium point are $(U_{+}, W_{+})=(4.20673, 0.106971)$ and starting from  (\ref{condHBF}), for fixed values of $\nu$, we find different values of the migration speed $c < c_\alpha^{HB}$, as $\alpha$ varies, such that an  oscillatory solution branch leaves from the Hopf bifurcation points.

\begin{enumerate}

\item{} For $\nu=460$, varying  $\alpha$ in the range $[0,1[$, we obtain the curve of migration speed $c_\alpha^{HB}$ given by (\ref{condHBF}) and reported in Fig. \ref{fig_c460}.  When $\alpha=0$,  $c_\alpha^{HB}=c^{HB}=11.3446$ according to   (\ref{cond}$_2$), see \cite{Sherratt_2011}.
When $\alpha\rightarrow 1$,  $c_\alpha^{HB}\rightarrow 0$, and according to \cite{Doelman_2013,Doelman_1998, Gandhi_2018} for $\alpha=1$ it has $c=0$.
In this context, we are not able to exactly quantify the very small values of $k_\alpha$ involved in $\tilde{\tilde{U}}_{+}$
and, consequently, in $c_\alpha^{HB}$, so we choose   
different values of $k_\alpha$ with the aim of showing the approximate behavior of the migration speed $c_\alpha^{HB}$ given by (\ref{condHB}), for which a Hopf bifurcation occurs.

We assume very small values of $k_\alpha$ 
preserving the behavior of the $c_\alpha^{HB}$ curve at the extremes values of $\alpha$ and we can choose $k_{\alpha}=\alpha(1-\alpha)/d^\alpha$. In Fig. \ref{fig_c460} we report $c_\alpha^{HB}$ with $d=10^4$ (blue line), $d=10^6$ (red line) and $d \rightarrow \infty $ (green line).
The choice to assume a very small $k_\alpha$ is confirmed by the three very close lines.

\begin{figure}[!ht]
\centering
\includegraphics[width=.45\textwidth]{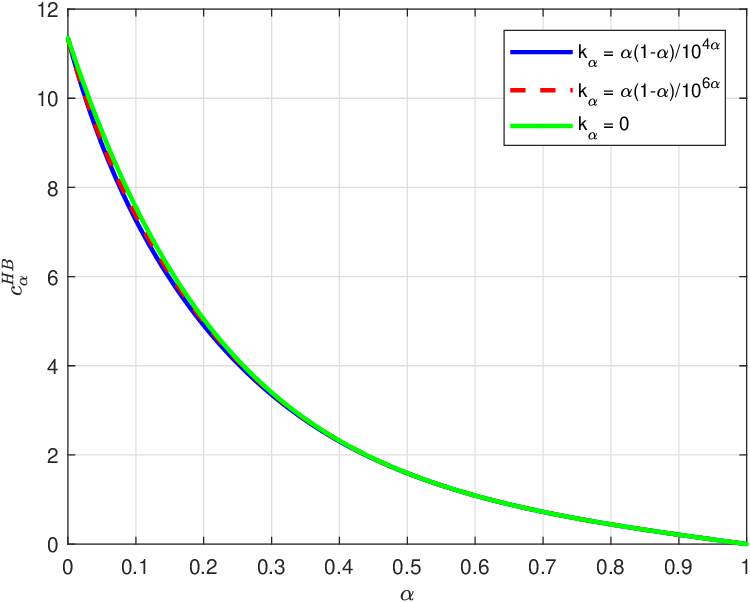} \qquad
\includegraphics[width=.45\textwidth]{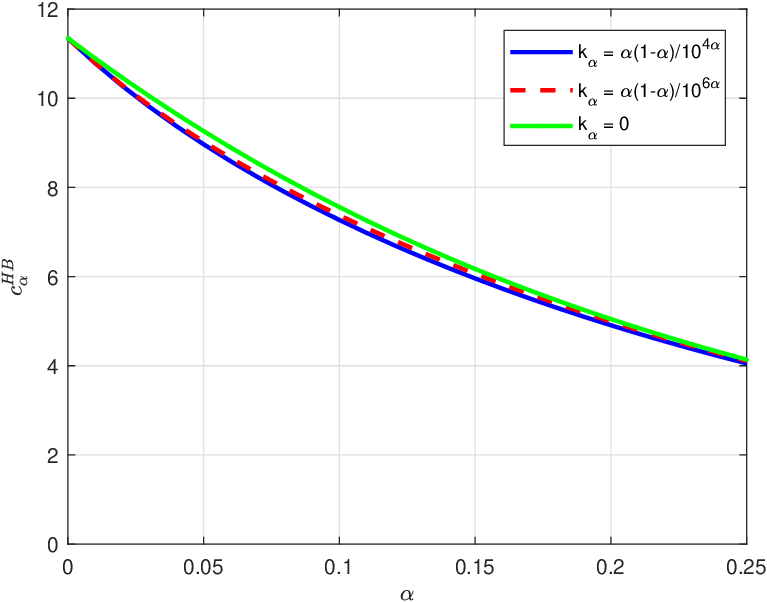} \caption{Migration speed $c^{HB}_\alpha$, depending on $\alpha$, given by (\ref{condHBF}) for $\nu=460$.  Right frame: zoom of the left frame.}
\label{fig_c460}
\end{figure}

\item{} For $\nu=380$, taking into account the considerations done in the previous case, for $\alpha=0$ it has  $c_\alpha^{HB}=c^{HB}=9.37162$ (see (\ref{cond})$_2$), and for $\alpha \rightarrow 1$ it has $c_\alpha^{HB} \rightarrow 0$, according to \cite{Doelman_2013,Doelman_1998,Gandhi_2018}.
Then we set, as in the previous case, $k_{\alpha}=\alpha(1-\alpha)/d^\alpha$. 
In Fig. \ref{fig_c380}, we report  $c_\alpha^{HB}$
 with $d=10^4$ (blue line), $d=10^6$ (red line) and $d \rightarrow \infty$ (green line).
As in the previous case, the choice to assume a very small $k_\alpha$ is confirmed by the three very close lines.

\begin{figure}[!h]
\centering
\includegraphics[width=.45\textwidth]{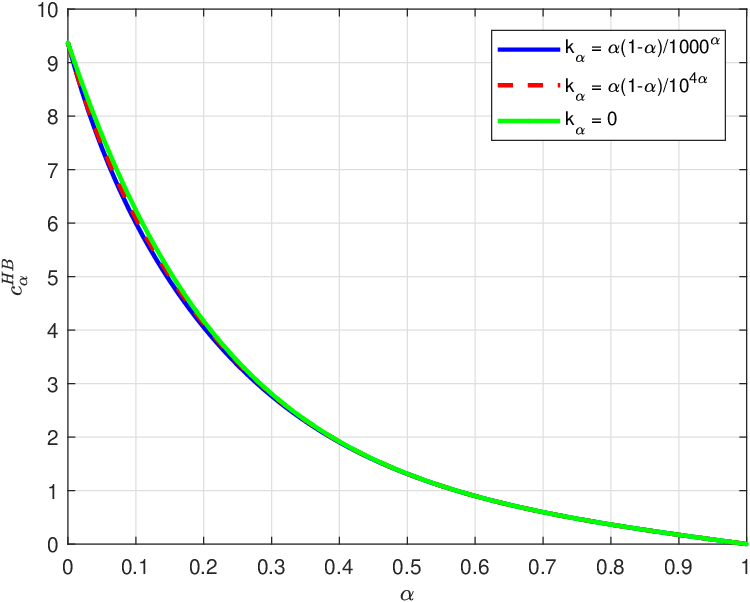} \qquad
\includegraphics[width=.45\textwidth]{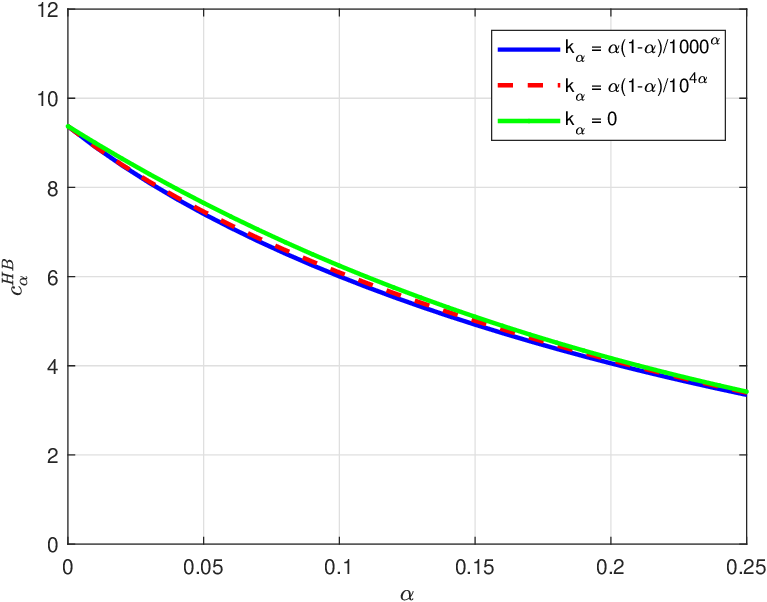}\caption{Migration speed $c^{HB}_\alpha$, depending on $\alpha$, given by (\ref{condHBF}) for $\nu=380$.  Right frame: zoom of the left frame.}
\label{fig_c380}
\end{figure}
 \end{enumerate}
Looking at the above figures (\ref{fig_c460}) and  (\ref{fig_c380}),  we remark that the variation of the behaviors of the migration speed curves, as the values of $k_\alpha$ approaches zero, is very small and we observe that the main variation occurs for values of $\alpha$ close to $0$ and becomes negligible for $\alpha$ approaching $1$.

\section{Numerical Method and Results}

The main goal of this paper is to study the new fractional model (\ref{eqF}) and to show that, as the parameter $\alpha$ varies,  it preserves the dynamics of the models (\ref{eqK}) and (\ref{eqGS}) and, then, the formation of the band vegetation patterns on sloped and no-sloped environments.  In this Section, we present the numerical method used for solving the fractional model (\ref{eqF}).
The numerical solutions are presented for different values of the fractional parameters $\alpha$. The comparison among numerical solutions, obtained by solving the models of integer order (\ref{eqK0}) and (\ref{eqGS0}), and the numerical solution of (\ref{eqF}) demonstrates the effectiveness and reliability of the new proposed fractional model (\ref{eqF}).
We use a one-step explicit numerical method arising from product integral rules, 
that is, the method based on the approximation of the integral formulation of the model under study. 
All numerical simulations are performed on Intel Core i7 by using Matlab software.

We rewrite the system (\ref{eqF}) in the following form
\begin{eqnarray*}
\begin{cases}
 U^{\prime \prime}=- c U^{\prime} +m U - U^2 W\\
 D_z^{\alpha+1}W= \displaystyle {\frac{1}{\nu}} \left( -c W^{\prime} +W +U^2 W- a  \right),
\end{cases}
\end{eqnarray*}
completed with suitable initial conditions.
As usual, we introduce the following assignments 
$$ ^1u(z) = U(z) , \qquad ^2u(z)  = W(z) , \qquad ^3u(z)  = U'(z) , \qquad ^4u(z)  = W'(z) ,$$
such that we obtain the system
\begin{eqnarray}\label{eqF1}
\begin{cases}
 ^1u' =  \ ^3u \\
 ^2u' = \ ^4u \\
 ^3u' = - c ^3u  + m \ ^1u - \ (^1u)^2 \ ^2u\\
 D_z^{\alpha} \ ^4u = \displaystyle {\frac{1}{\nu}} \left( -c \ ^4u + \ ^2u  + \  (^1u)^2 \ ^2u - a  \right).
\end{cases}
\end{eqnarray}

In order to solve the fractional model (\ref{eqF1}), we set a positive integer number $J$ and  define a uniform computational grid of $J+1$ grid-points, namely $z_j$, with  $z_j = z_0 + j \Delta z$,
for $j=0,\cdots, J$ and integration step size $\Delta z $.
Then, at each the grid--points $z_j$, we define with $U_j$ and  $W_j$ the numerical approximations of the exact solutions $U(z_j)$ and $W(z_j)$,  the density of plant and water, respectively.

Starting from the fourth equation of the above system, we introduce its equivalent Volterra integral formulation 
\begin{equation}\label{eqF_V}
^4u(z) = \ ^4u_0 + \frac{1}{\Gamma(\alpha)} \int^z_{z_0}  f(s, \ ^4u(s)) (z-s)^{\alpha -1}d s ,
\end{equation}
where 
$$ f(z, \ ^4u(z))  = \displaystyle {\frac{1}{\nu}} \left( -c \ ^4u + \ ^2u  + \  (^1u)^2 \ ^2u - a  \right) .$$
For the problem under study, the equivalent Volterra integral equation at $z=z_{j+1}$ reduces to 
\begin{eqnarray}\label{Int_V}
^4u(z_{j+1}) &=& ^4u_0 + \frac{1}{\Gamma(\alpha)} \int^{z_{j+1}}_{z_0} f(s,\ ^4u(s)) (z_{j+1}-s)^{\alpha -1}d s \nonumber \\
&=& ^4u_0 + \frac{1}{\Gamma(\alpha)} \sum_{k=0}^j \int^{z_{k+1}}_{z_k} f(s,\ ^4u(s)) (z_{j+1}-s)^{\alpha -1}d s  \ .
\end{eqnarray}
Now, in each sub-interval $[z_{k}, z_{k+1}]$ we can substitute the function $f (z, ^4u(z))$ with an interpolation polynomial, so that the resulting integrals can be exactly evaluated. In order to compute $ ^4u(z_{j+1})$, to approximate the integral on the right-hand side of the (\ref{Int_V}), we use the rectangular explicit rule, obtaining the following explicit formula
\begin{eqnarray}\label{FEE}
^4u_{j+1} = ^4u_0 + \frac{1}{\Gamma(\alpha)} \sum_{k=0}^j a_{k} f (z_{k}, \ ^4u_{k}) \qquad j=0,\cdots,J-1 \,
\end{eqnarray}
with
\begin{eqnarray*}
 a_{k} = \int^{z_{k+1}}_{z_k} (z_{j+1}-s)^{\alpha -1}d s  = \frac{1}{\alpha} \left ( (z_{j+1} - z_{k})^\alpha  - (z_{j+1} - z_{k+1})^\alpha \right) \ .
\end{eqnarray*}
Finally, operating in the same way for the first three equations, we obtain the following explicit method (PI$_1$ Ex) with an order of accuracy equal to one
\begin{eqnarray}\label{met}
\begin{cases}
 ^1u_{j+1}  = \ ^1u_{j}  + \Delta z \ ^3u_j \\
 ^2u_{j+1}  = \ ^2u_{j}  + \Delta z \ ^4u_j \\
 ^3u_{j+1}  = \ ^3u_{j}  + \Delta z \left(- c \ ^3u_j  + m \ ^1u_j - \ (^1u_j)^2 \ ^2u_j \right) \\
 ^4u_{j+1} = \ ^4u_0 +   \displaystyle {\frac{1}{\nu \ \Gamma(\alpha)}} \sum_{k=0}^j a_{k} \left(- c \ ^4u_k + \ ^2u_k  + \  (^1u_k)^2 \ ^2u_k - a  \right).
\end{cases}
\end{eqnarray}

In the following, we report some numerical applications for $a=2$ and $m=0.45$, cases introduced in Sect. 3.1. We solve the system (\ref{eqF}) with the following initial conditions
$$ ^1u _0 = 3.87016, \qquad  ^2u _0 = \frac{m}{^1u_0}, \qquad  ^3u _0 = 0, \qquad  ^4u _0 = 0, $$
on computational domain $[-100,500]$ with $J=10000$ and $\Delta z = 0.06$. The initial conditions $^1u _0$ and $^2u _0$ are chosen as perturbations of the equilibrium point $(U_+,W_+)=(4.20673, 0.106971)$, obtained by (\ref{equ_points}).  Moreover, the migration speed $ c$ is set such that $c =c_{\alpha} <  c_{\alpha}^{HB}$ to obtain oscillatory solutions.

\bigskip

{\bf Remark 1. }
For the problem under study, as far as the performed numerical simulations of the fractional model, we need to use a large number of nodes to obtain highly accurate solutions and then, in order to maintain the computational cost not too high, we propose an explicit first-order method.

\subsection{Numerical applications}

To validate the theoretical results, we present some applications of interest concerning the vegetation pattern formation obtained by numerically solving the proposed fractional model.
By the numerical solutions, we show that vegetation pattern formation arises when the migration speed $c= c_\alpha$ assumes smaller values than the migration speed  $c_\alpha^{HB}$, according to the theory performed by Sherratt  \cite{Sherratt_2011} for the KL model. It is important to note that, as the parameter $\alpha$ varies, the new proposed fractional model preserves the solution pattern formation { and guarantees the dynamics}.
 
\subsection*{Test 1}
\begin{figure}[!h]
\centering
\includegraphics[width=.30\textwidth]{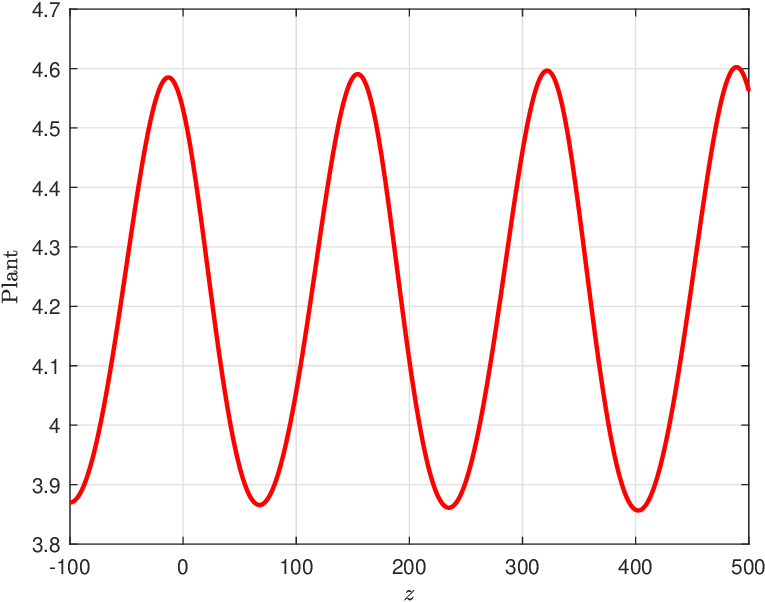} \quad
\includegraphics[width=.30\textwidth]{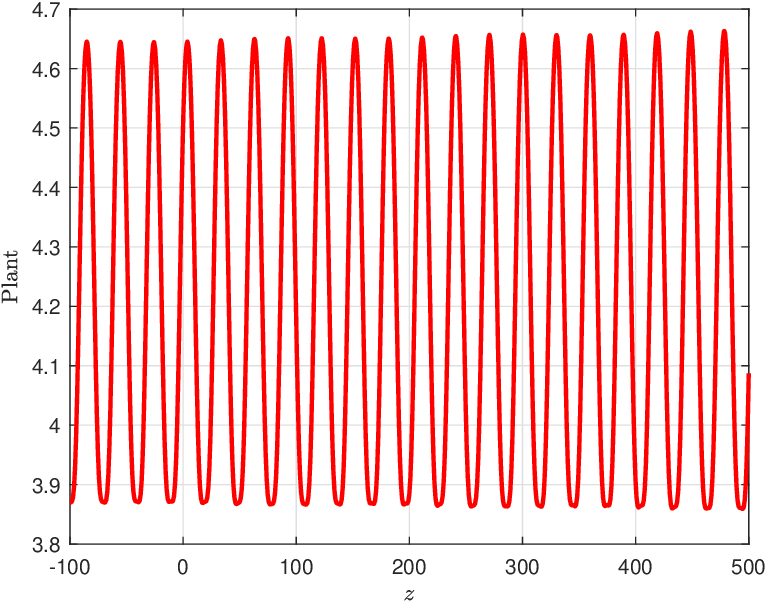} \\
\includegraphics[width=.30\textwidth]{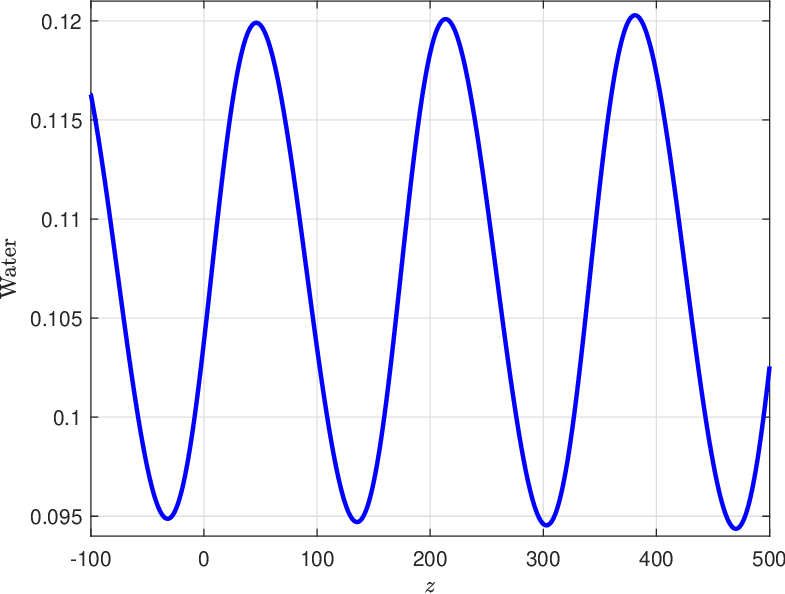} \quad
\includegraphics[width=.30\textwidth]{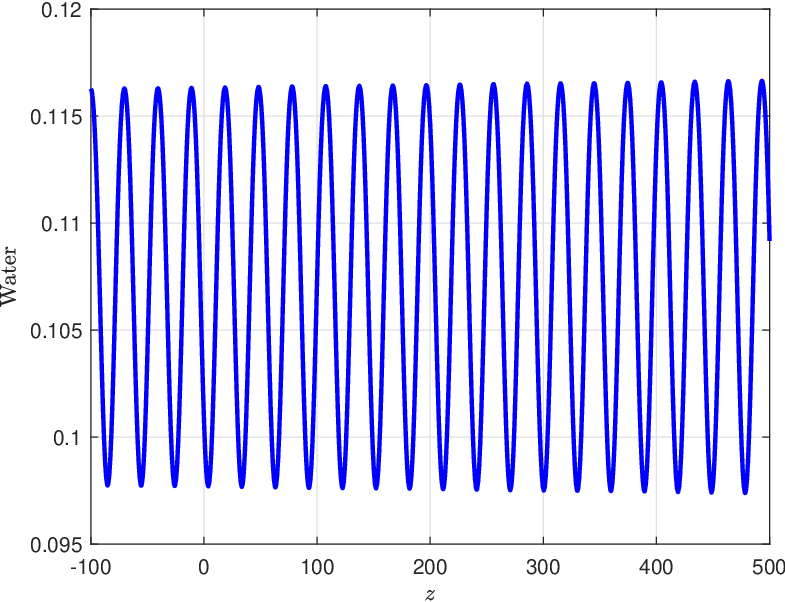} 
\caption{Test 1. Numerical solutions of the reduced models: Top frames: numerical solution $U_j$. 
Bottom frames: numerical solution $W_j$.
Left frames: the solution $U_j$ of the reduced KL model. 
Right frames: the solution $W_j$ of reduced KL--GS model.}
\label{fig1_0}
\end{figure}

\begin{figure}[!h]
\centering
\includegraphics[width=.30\textwidth]{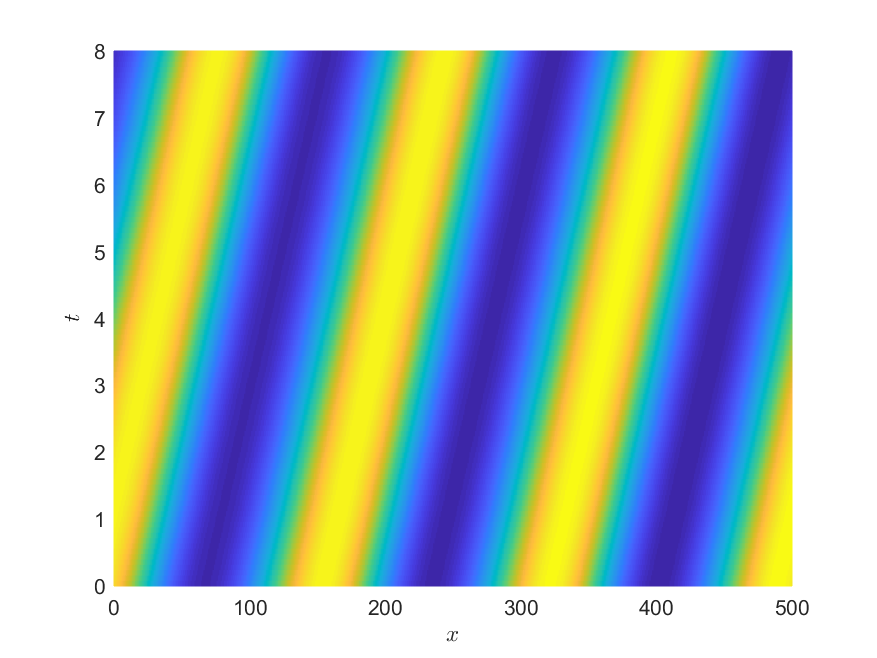}  \quad
\includegraphics[width=.30\textwidth]{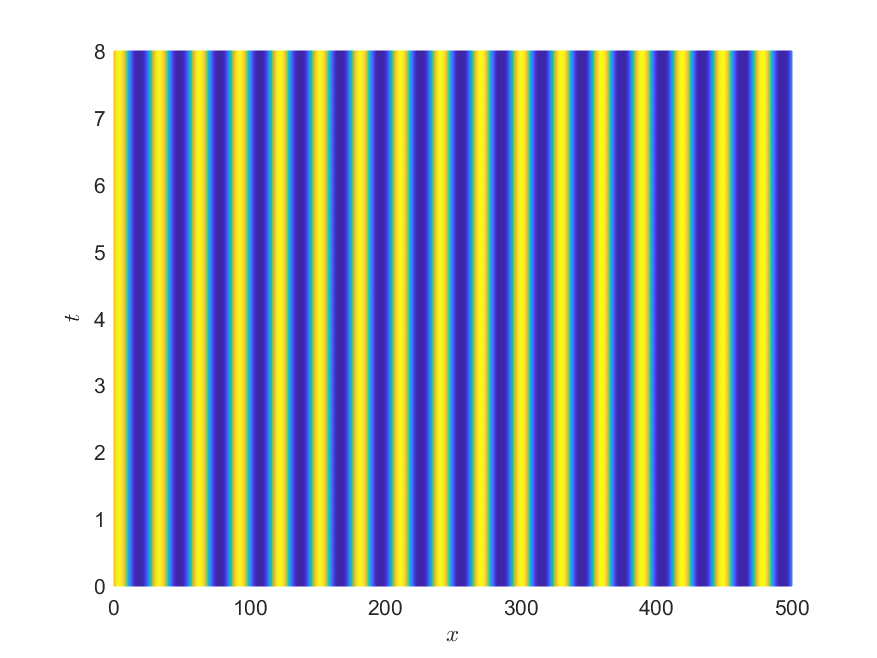} \\
\includegraphics[width=.30\textwidth]{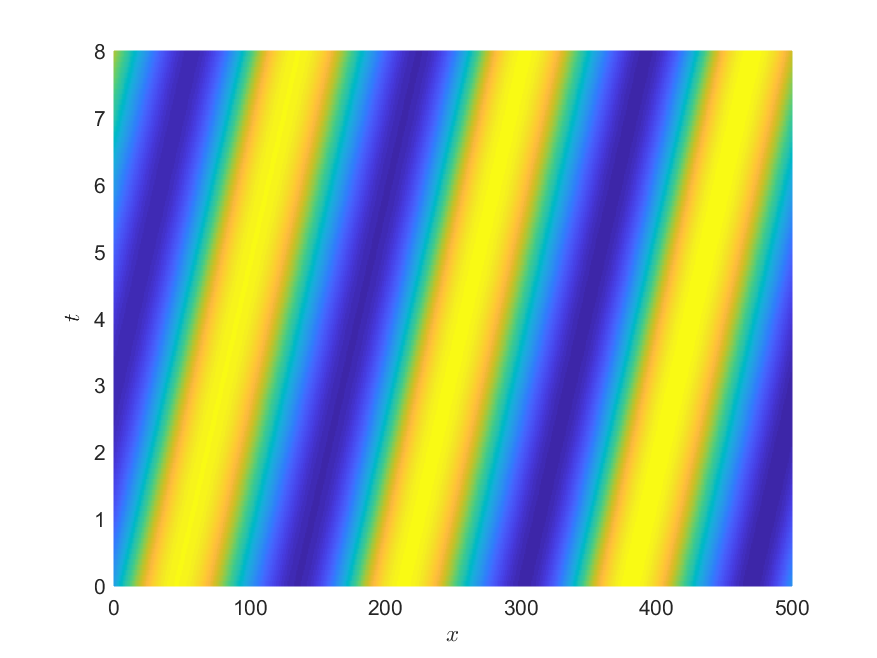}  \quad
\includegraphics[width=.30\textwidth]{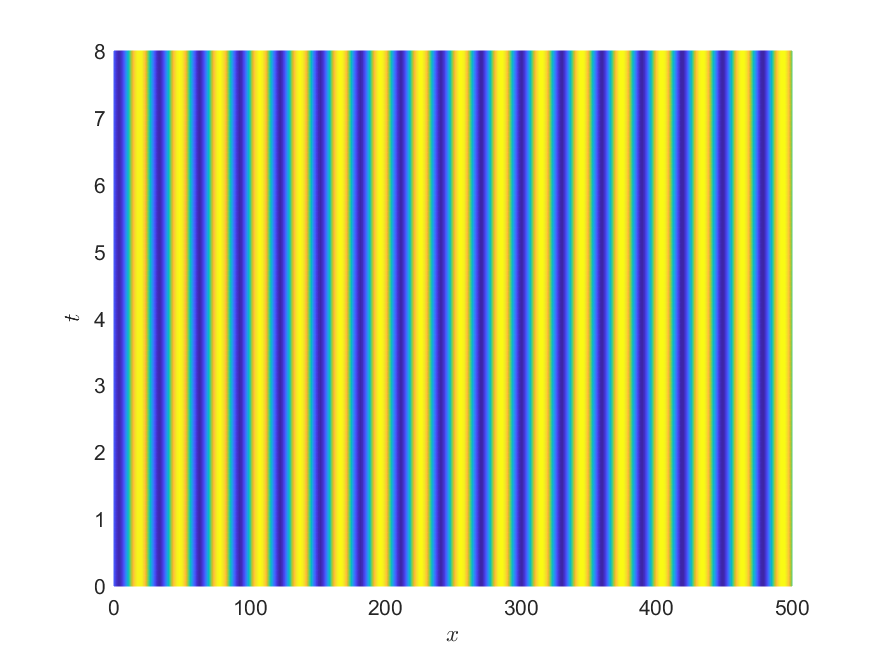} 
\caption{Test 1. Numerical solutions: Top frames: numerical solution 
$u_j^n$. 
Bottom frames: numerical solution $w_j^n$. 
Left frames: numerical solutions $u_j^n$ and $w_j^n$ of KL model. 
Right frames: numerical solutions $u_j^n$ and $w_j^n$ of KL--GS model.}
\label{fig1_00}
\end{figure}

In this first test, we set the parameters, involved in the models as follows 
$$a   = 2, \quad m   = 0.45, \qquad \nu  = 460.$$
In Figure \ref{fig1_0}, we report the numerical solutions of the reduced Klausmeier  (\ref{eqK0}) model with $c=11.29$ and reduced Klausmeier--Gray--Scott model (\ref{eqGS0}) ($c=0$), obtained by the classical first-order explicit Euler method.
The vegetation patterning process on sloped terrains described by the KL model is shown in the top frames of Figure \ref{fig1_00}.
The vegetation patterning process on no--sloped, flat,  terrains described by the KL--GS model is shown in the bottom frames of Figure \ref{fig1_00}. The solutions $u_j^n$ and $w_j^n$ are reconstructed by using the cubic spline interpolation to evaluate them at any point of the computational domain, taking into account that $z=x- c t$ with final time $t=8$.

In the following for solving the fractional model, we set $ c=c_\alpha$,  depending on the parameter $\alpha$, as reported in Table \ref{tab1}. The values, reported in Table \ref{tab1}, lead to the pattern formation.
For $\alpha =0$, in the KL model, we find the migration speed  $c =11.29 < c^{HB} = 11.3446$,  the value found to obtain periodic solutions with a constant wavelength, following Sherratt  \cite{Sherratt_2011}.
In the same way, we find the values of the velocity $c_\alpha < c_\alpha^{HB}$, such as obtaining oscillatory solutions with a wavelength for each parameter value $\alpha$. Moreover, for $\alpha \rightarrow 1$ we have $c_\alpha \rightarrow 0$. 

\small{\begin{table}[!h]
\begin{tabular}{|c|c|c|c|c|c|c|c|c|c|c|c|} 
\hline
$\alpha$ &  0.0 & 0.1 & 0.2 & 0.3 & 0.4 & 0.5 & 0.6 & 0.7 & 0.8 & 0.9 & $\rightarrow$  1 \\
\hline 
$c_\alpha$ & 11.29 & 7.155 & 4.753& 3.238& 2.235& 1.539& 1.040 &  0.668 & 0.386 & 0.149  &  $\rightarrow$ 0  \\
   \hline 
\end{tabular}
\caption{Values of the velocity $c_\alpha$ depending on the fractional parameter $\alpha$.}  
\label{tab1}
\end{table}}

\begin{figure}[!h]
\centering
\includegraphics[width=.35\textwidth]{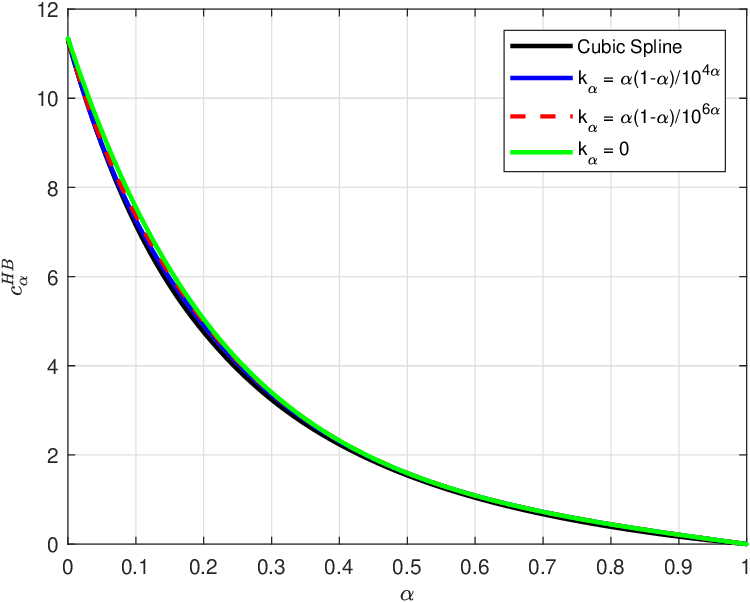} \qquad
\includegraphics[width=.35\textwidth]{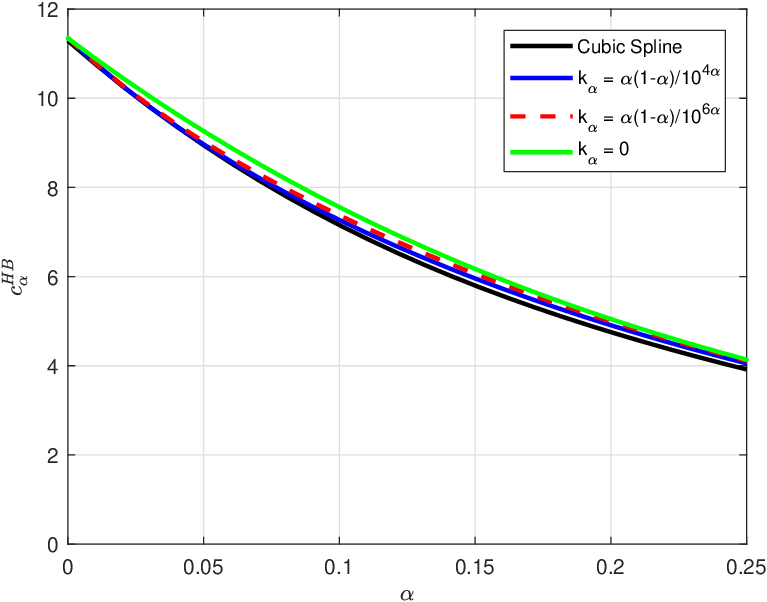}\caption{Migration speed $c^{HB}_\alpha$, depending on $\alpha$, given by (\ref{condHBF}) for $\nu=460$. Right frame: zoom of the left frame.}
\label{fig1_c460}
\end{figure}

The values ($\alpha$, $c_\alpha$), reported in Table \ref{tab1}, are used as interpolation nodes to build the interpolation cubic spline $p(\alpha)$ and then to obtain the values of $c_\alpha$, $\forall \alpha $ with $0<\alpha<1$. In Fig. \ref{fig1_c460}, we show the migration speeds $c_\alpha^{HB}$,  with $d=10^4$ (blu line), $d=10^6$ (red line) and $d \rightarrow \infty$ (green line), and $p(\alpha)$ obtained by the interpolation with the cubic spline (black line). The obtained results confirm that the assumption of small and positive values of $k_\alpha$ is valid.

\begin{figure}[!h]
\centering
\begin{subfigure}
\centering
\includegraphics[width=.245\textwidth]{Test1_P_KL_2d.eps} \quad
\includegraphics[width=.245\textwidth]{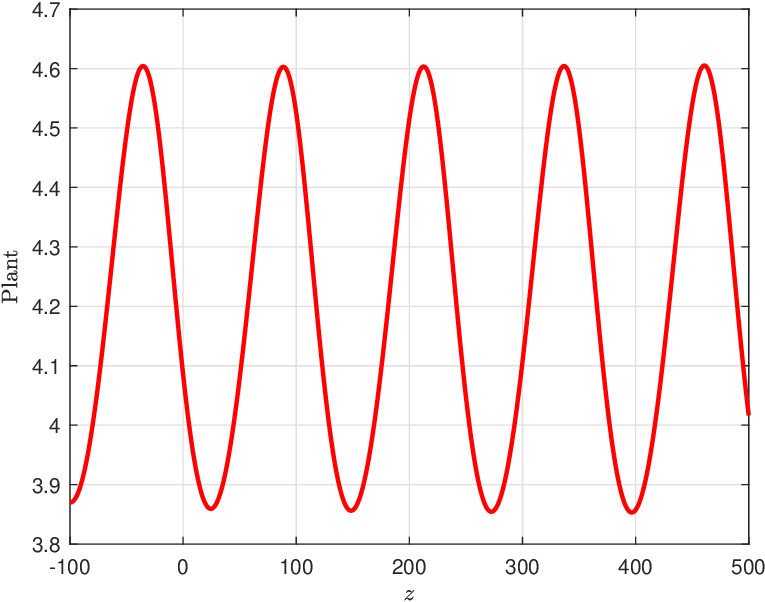}\quad 
\includegraphics[width=.245\textwidth]{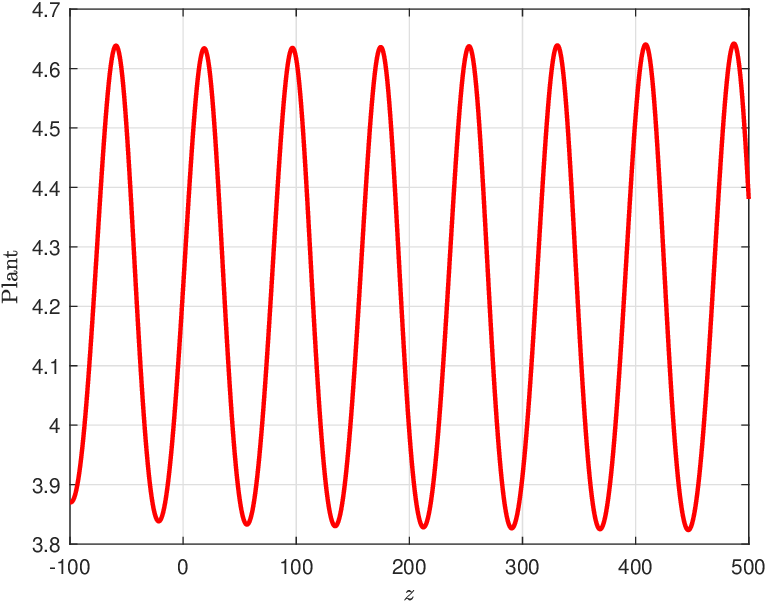} \\
\includegraphics[width=.245\textwidth]{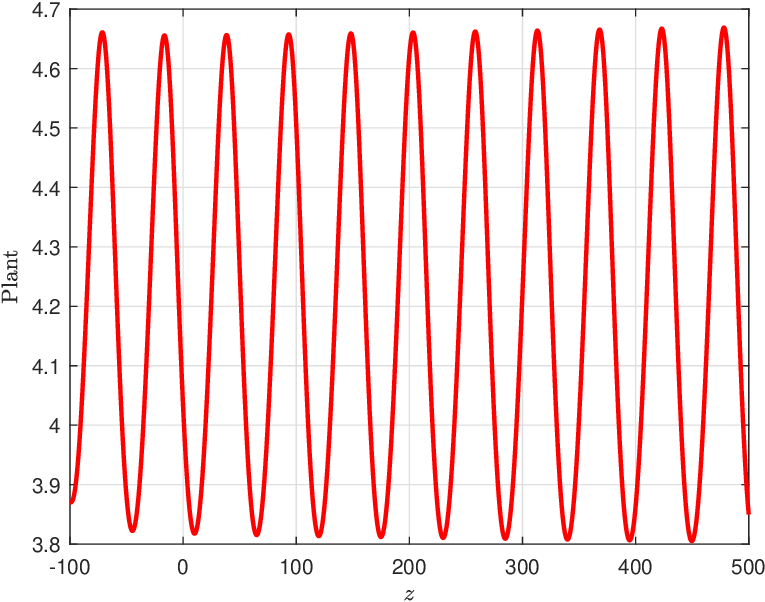} \quad
\includegraphics[width=.245\textwidth]{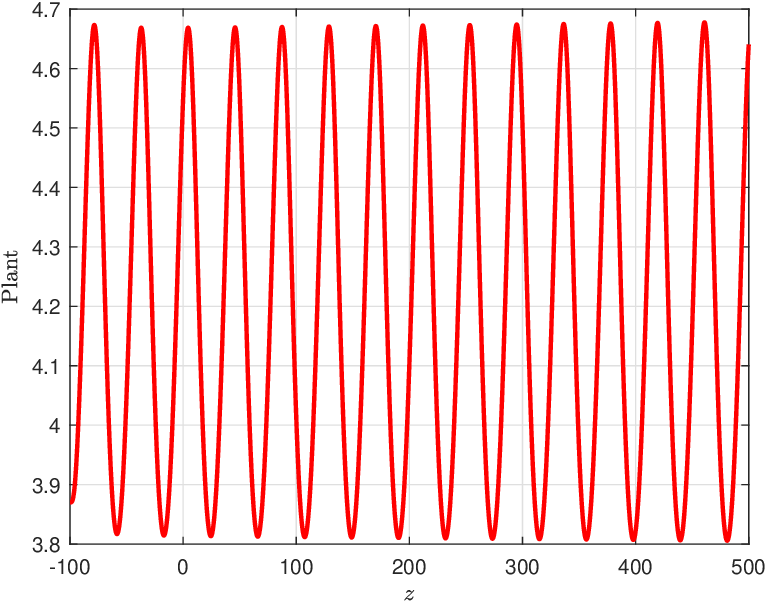} \quad
\includegraphics[width=.245\textwidth]{Test1_P_GS_2d.eps}
\caption{Test 1. Numerical solutions $U_j$ of the FM related to the concentration of the plant $U(z_j)$ for different values of $\alpha$. From top to bottom: $\alpha= 10^{-8}$, $\alpha=0.1$, $\alpha =0.3$, $\alpha=0.5$, $\alpha =0.7$ and $\alpha = .999999$.}
\label{fig1_1}
\end{subfigure}
\bigskip
\begin{subfigure}
\centering
\includegraphics[width=.245\textwidth]{Test1_W_KL_2d.eps} \quad
\includegraphics[width=.245\textwidth]{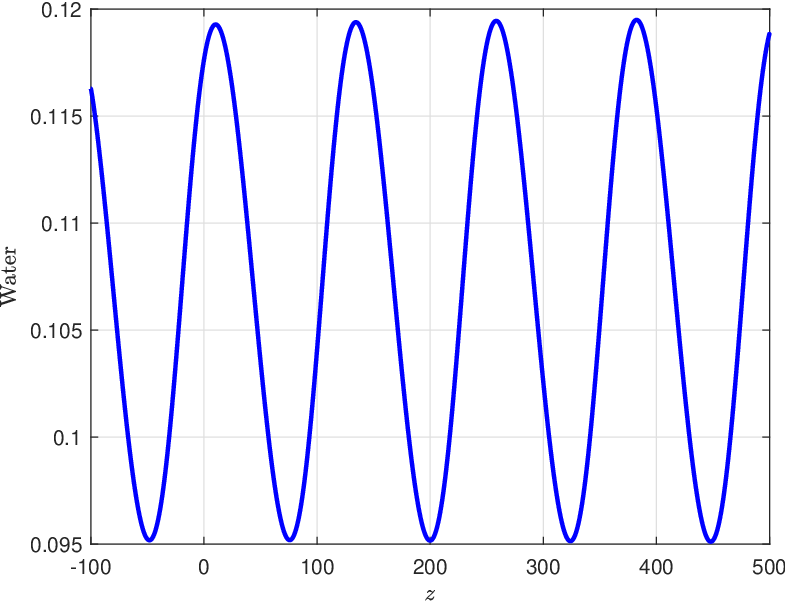} \quad
\includegraphics[width=.245\textwidth]{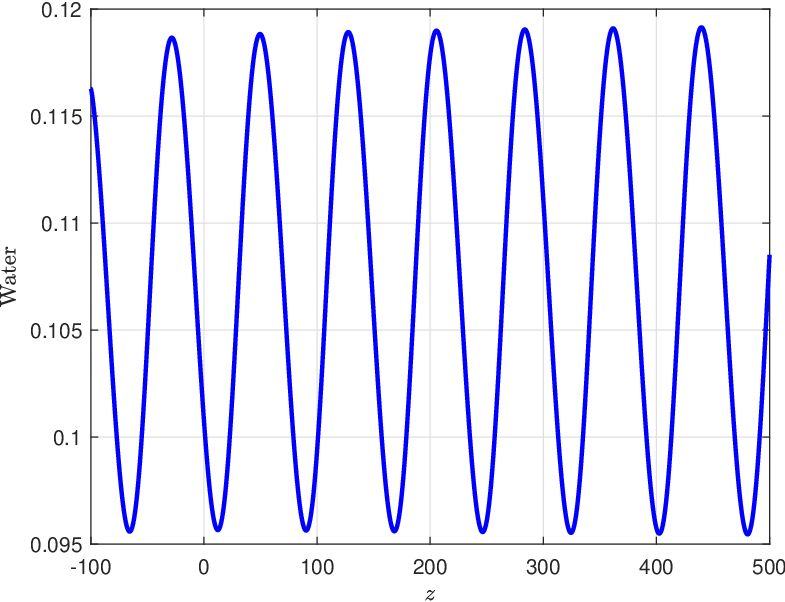} \\
\includegraphics[width=.245\textwidth]{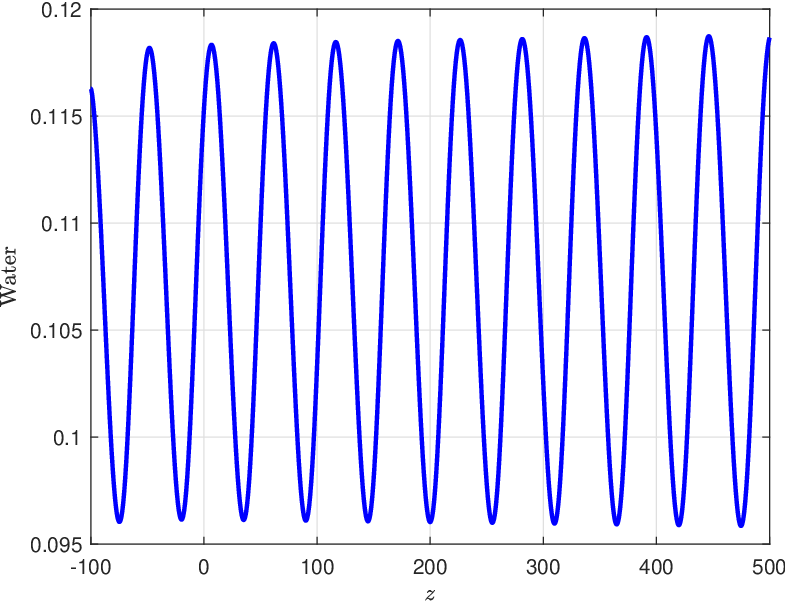} \quad
\includegraphics[width=.245\textwidth]{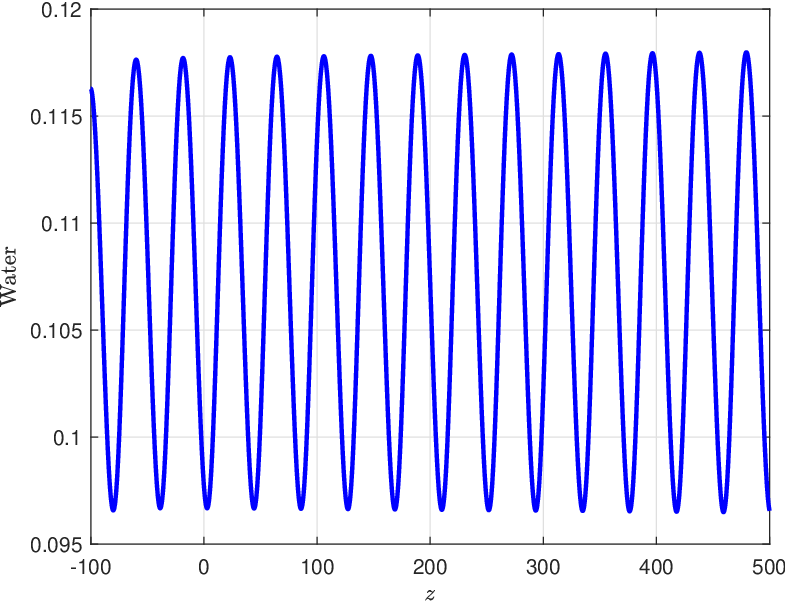} \quad
\includegraphics[width=.245\textwidth]{Test1_W_GS_2d.eps} 
\caption{Test 1. Numerical solutions $W_j$ of the FM related to the concentration of the water $W(z_j)$ for different values of $\alpha$. From top to bottom: $\alpha= 10^{-8}$, $\alpha=0.1$, $\alpha =0.3$, $\alpha=0.5$, $\alpha =0.7$ and $\alpha = .999999$.}
\label{fig1_2}
\end{subfigure}
\end{figure}

In Figures \ref{fig1_1} and \ref{fig1_2}, we report the numerical solutions $U_j$ and $W_j$ of the fractional model (\ref{eqF}) obtained for different values of $\alpha$, with $0< \alpha<1$ and with the corresponding migration speed $c_\alpha$ reported in Tab.\ref{tab1}.

Note that, as the value of $\alpha $ increases,  the solution maintains the oscillatory behaviour. The maximum and minimum values of the solutions vary slightly as $\alpha$ varies.
As the speed of migration decreases the wavenumber (the spatial frequency of a wave in the considered domain) increases and the wavelength (the distance between two consecutive points that are in the same phase) decreases.

\begin{figure}[!h]
\centering
\begin{subfigure}
\centering
\includegraphics[width=.25\textwidth]{Test1_P_KL.eps}  \quad
\includegraphics[width=.25\textwidth]{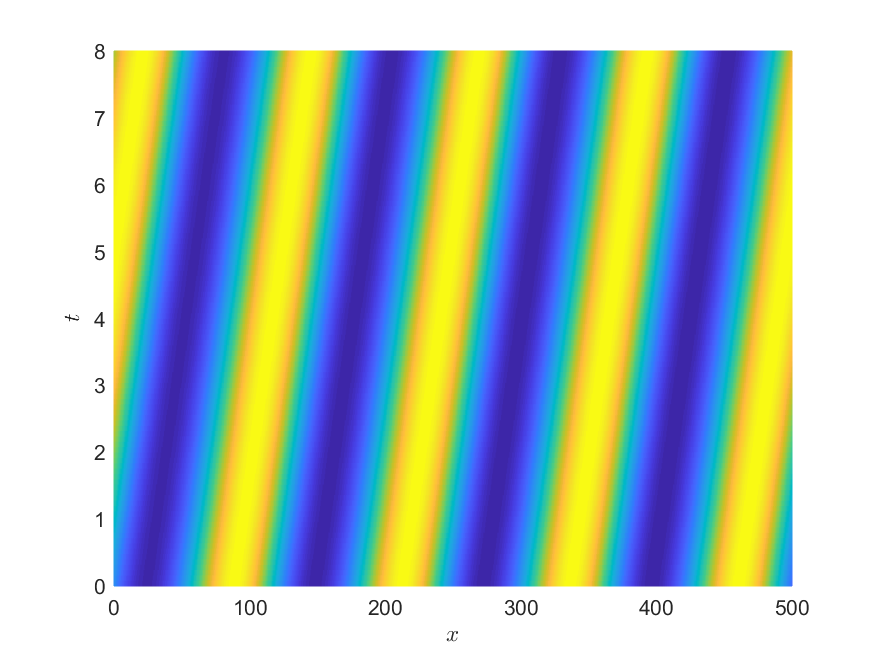} \quad
\includegraphics[width=.25\textwidth]{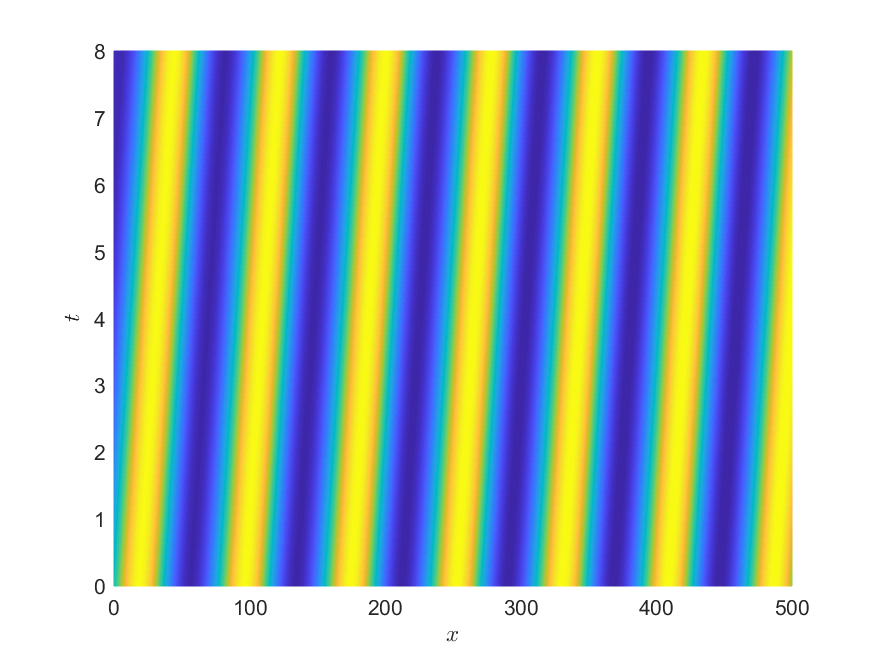} \\
\includegraphics[width=.25\textwidth]{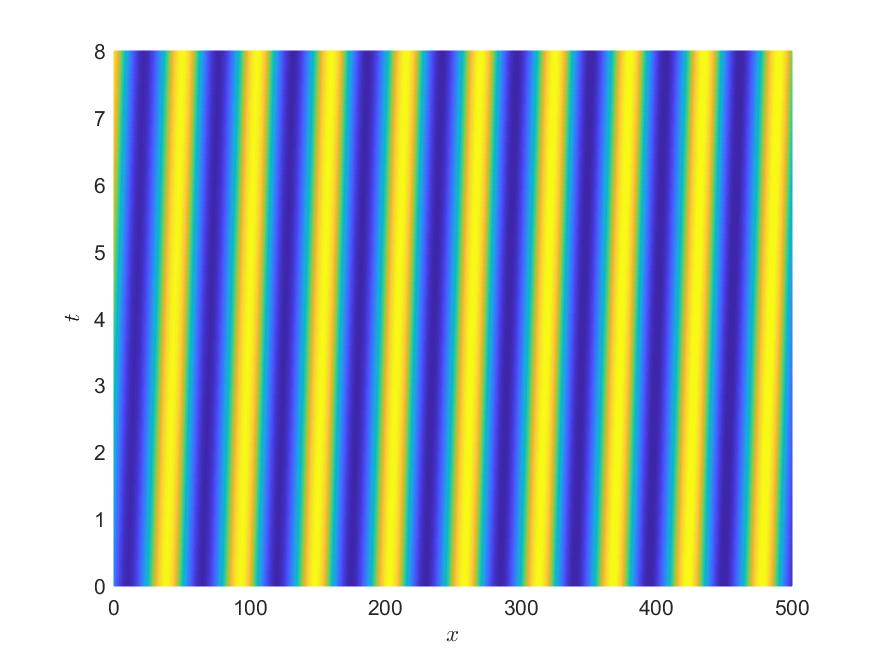} \quad
\includegraphics[width=.25\textwidth]{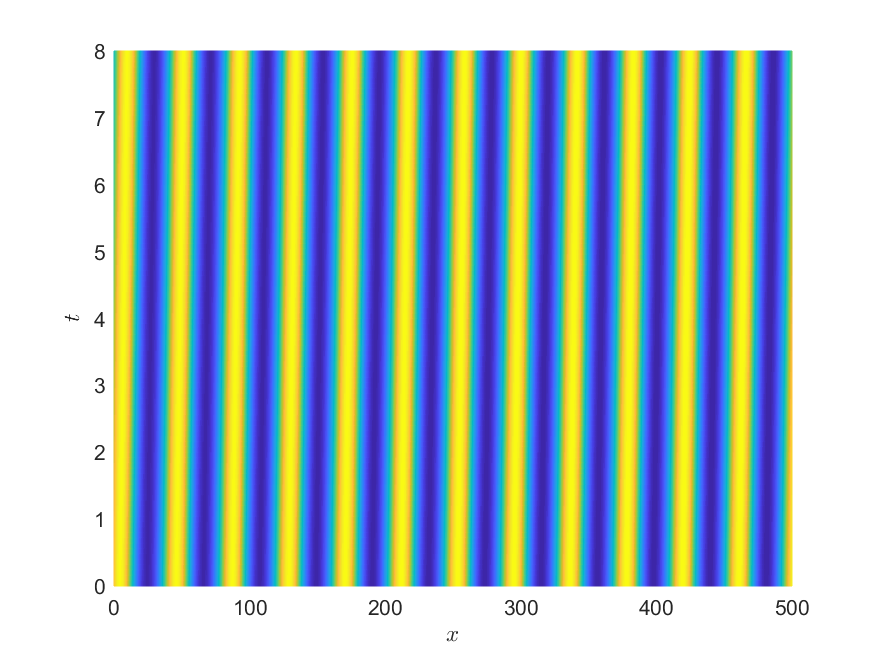} \quad
\includegraphics[width=.25\textwidth]{Test1_P_GS.eps} 
\caption{Test 1. Numerical solutions of the concentration of plants $u(t,x)$ at final time $t=8$. From top to bottom: with $\alpha= 10^{-8}$, $\alpha=0.1$,  $\alpha =0.3$, $\alpha=0.5$, $\alpha =0.7$ and $\alpha = .999999$.}
\label{fig1_3}
\end{subfigure}
\bigskip
\begin{subfigure}
\centering
\includegraphics[width=.25\textwidth]{Test1_W_KL.eps}  \quad
\includegraphics[width=.25\textwidth]{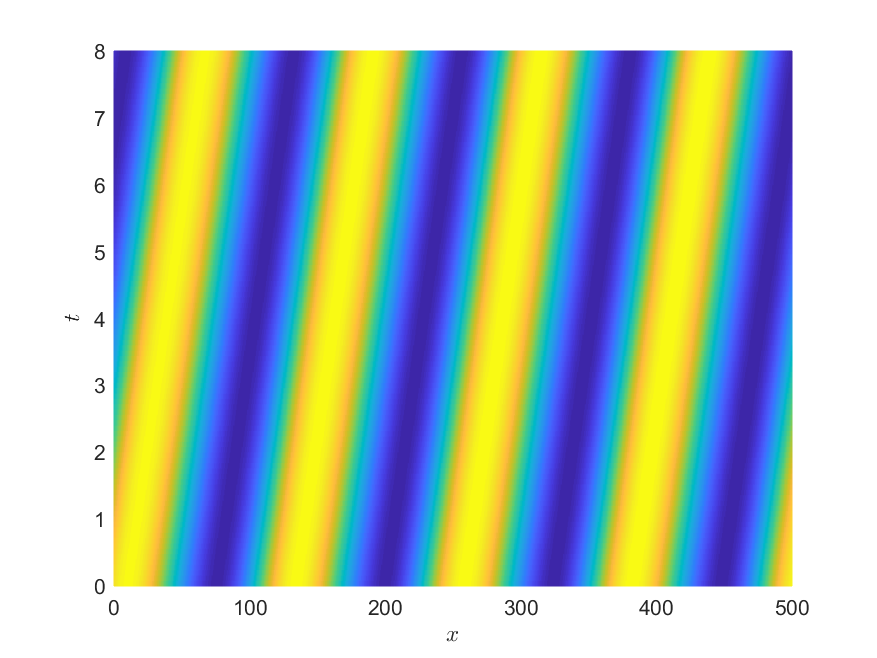} \quad
\includegraphics[width=.25\textwidth]{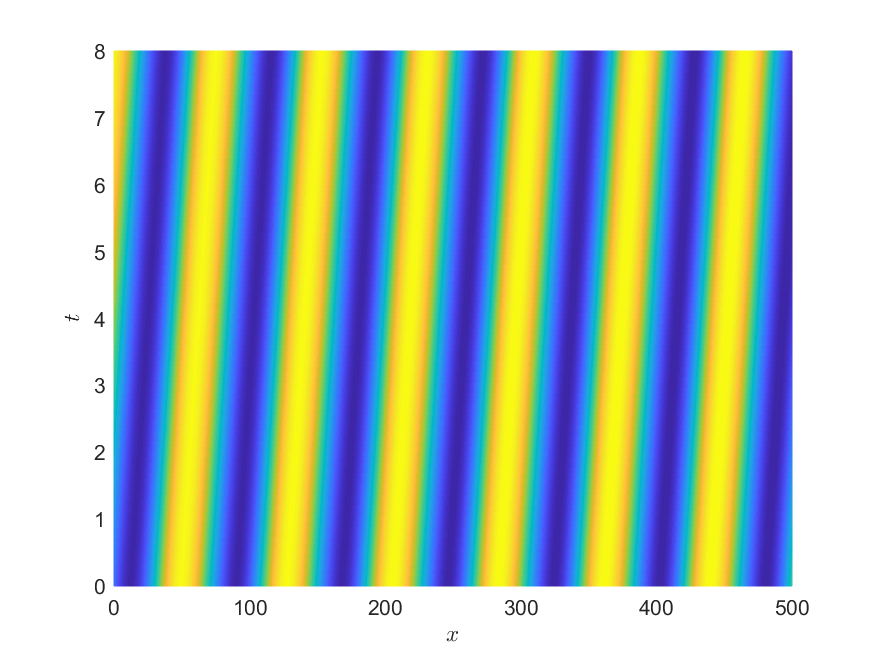} \\
\includegraphics[width=.25\textwidth]{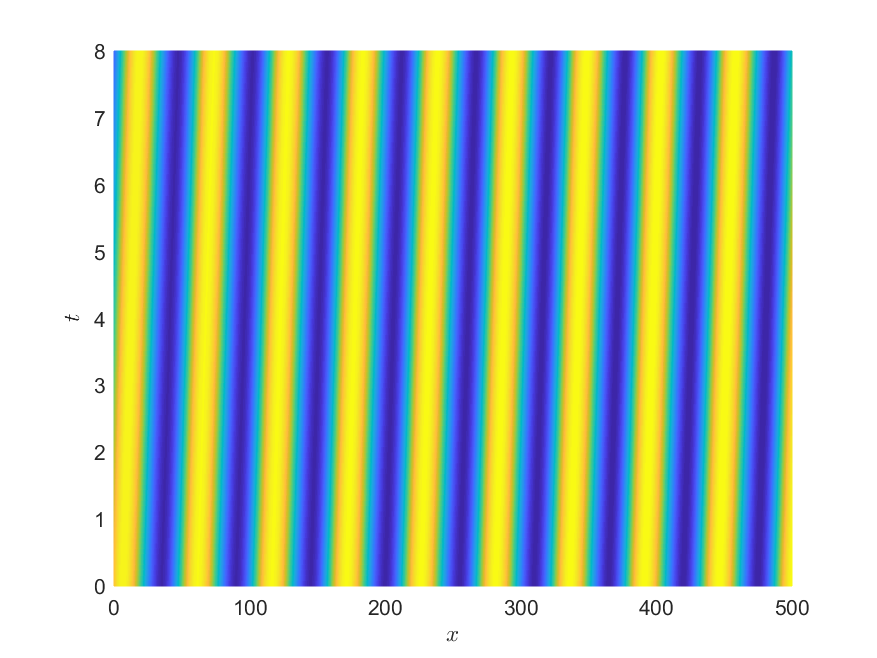} \quad
\includegraphics[width=.25\textwidth]{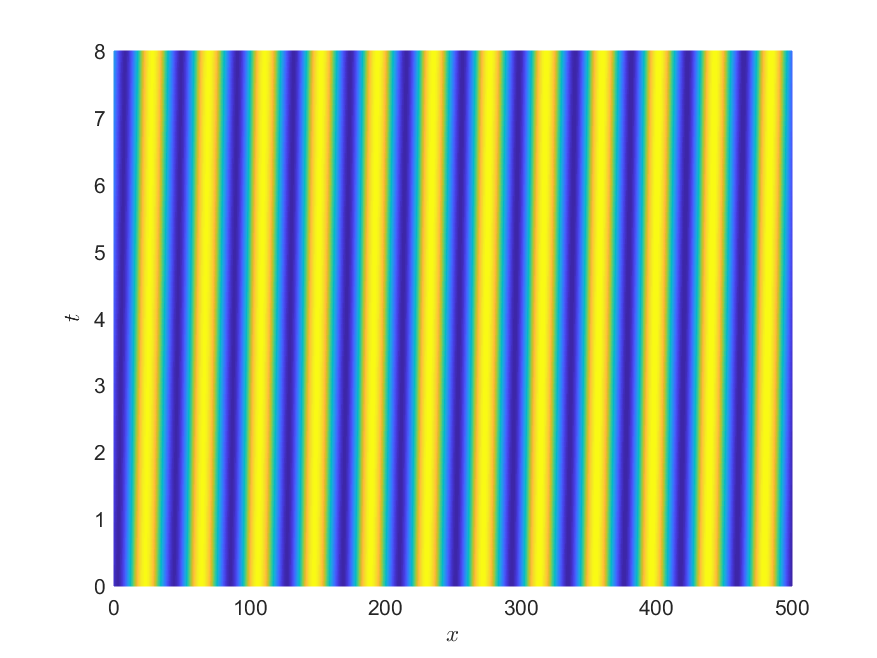} \quad
\includegraphics[width=.25\textwidth]{Test1_W_GS.eps} 
\caption{Test 1. Numerical solutions of the concentration of water  $w(t,x)$ at final time $t=8$. From top to bottom: $\alpha= 10^{-8}$, $\alpha=0.1$, $\alpha =0.3$, $\alpha=0.5$, $\alpha =0.7$ and $\alpha = .999999$.}
\label{fig1_4}
\end{subfigure}
\end{figure}

In Figures \ref{fig1_3} and \ref{fig1_4}, 
we report the numerical solutions $u_j^n$ and $w_j^n$ obtained for different values of $\alpha$, taking into account that $z=x- c_\alpha t$ with final time $t=8$.
The solutions $u_j^n$ and $w_j^n$ are reconstructed by using the cubic spline interpolation to evaluate them at any point of the computational domain.
The numerical results show the vegetation pattern migration and how they exhibit different trajectories due to the different values of the migration speed that depend on $\alpha$ and, then, by the slope of the domain. { So that, we can observe the migration of the plants in uphill direction}. 
On gentle slopes, the solutions are generally made up of bands of vegetation parallel to the level curves, separated by bands of bare ground.
As shown in Figures \ref{fig1_3} and \ref{fig1_4}, we report space-time plots of the location of the water and vegetation patterns. The trajectories of the solutions reveal the anomalous transport of the concentrations: the direction of the migration is towards increasing $x$, corresponding to the uphill direction.
The patterns move at a constant migration speed for each value of $\alpha$ in the positive direction corresponding to the uphill migration.

We remark that, for $\alpha=0$, the fractional derivative is the first-order derivative that describes a classical advection process and for
$\alpha=1$, the fractional derivative is the second-order derivative that describes a classical diffusion process. Note that, for $\alpha= 10^{-8}$, the obtained solution is very similar to the numerical solution of the KL model, where the classical advection process arises.
Moreover, for $\alpha = .999999$, the water advection tends to be irrelevant and the obtained solution is very similar to the numerical solution of the KL--GS model, where the classical diffusion process arises.

The fractional formulation of the mathematical model allows to get the  oscillatory behavior of the solutions and, therefore, ensures the vegetation pattern formation and migration in arid and semi-arid environments with any slope.

\begin{figure}[!h]
\centering
\includegraphics[width=.34\textwidth]{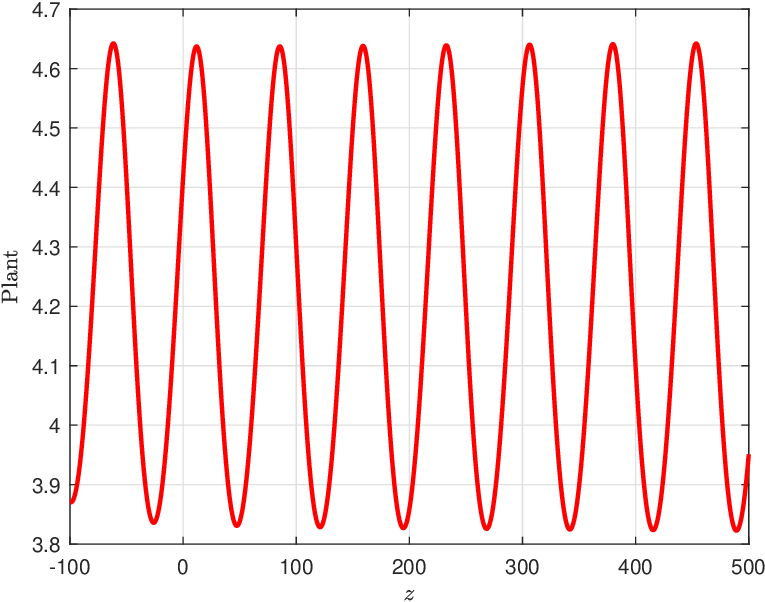} \qquad
\includegraphics[width=.38\textwidth]{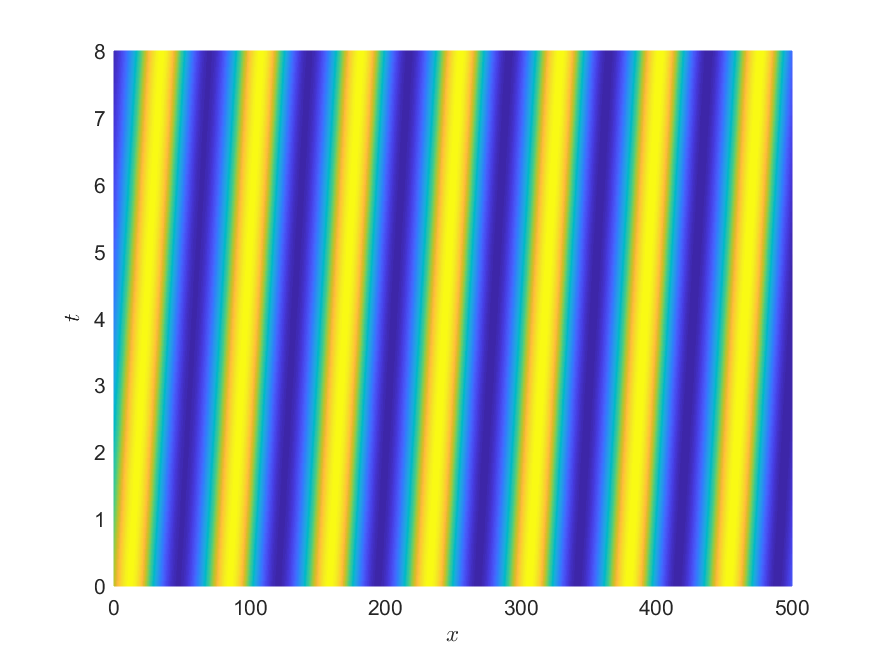} \\
\includegraphics[width=.34\textwidth]{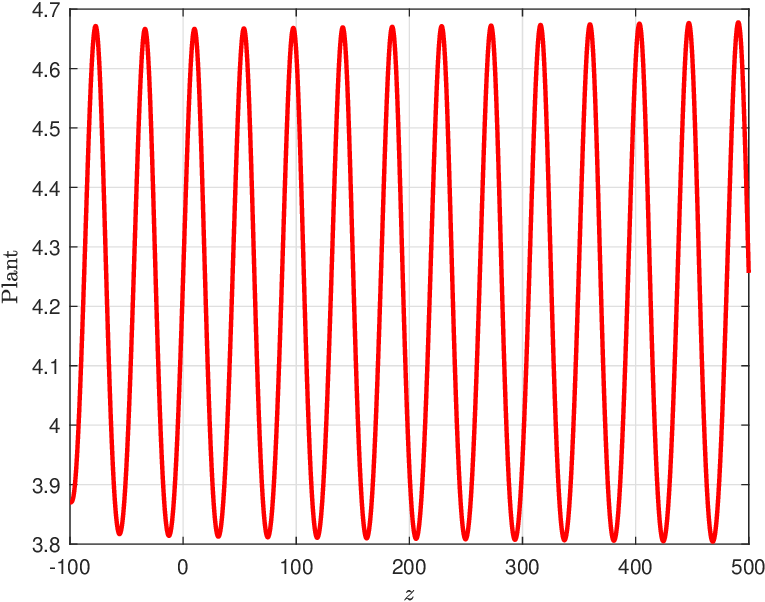} \qquad
\includegraphics[width=.38\textwidth]{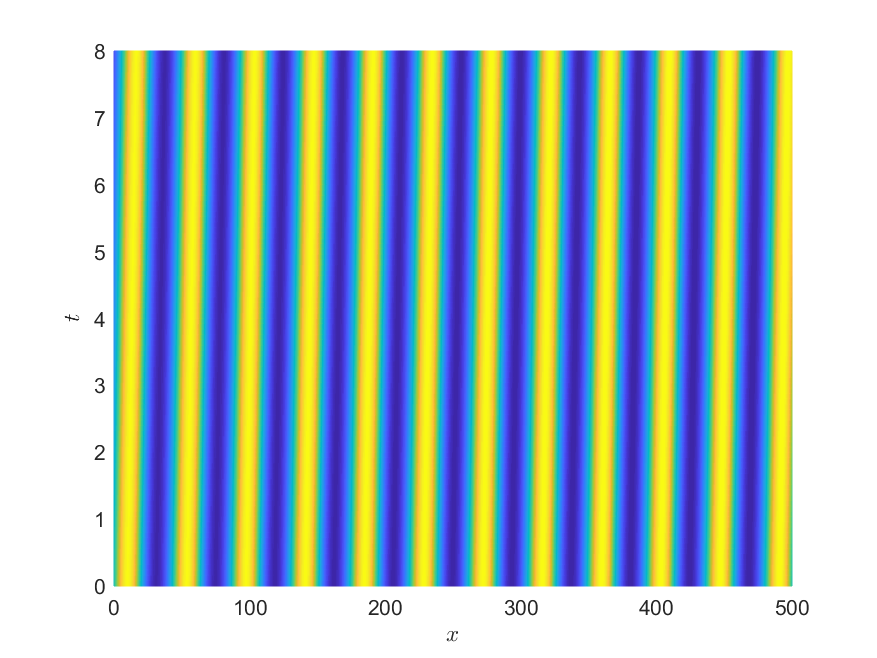} 
\caption{Test 1. The numerical solutions $U_j$ of $U(z)$ and $u_j^n$ of $u(x,t)$ obtained for $a=2$, $m=0.45$, $ \nu = 460$. Top: $\alpha=0.33$ and $c_\alpha=  2.895$. Bottom: $\alpha=0.66$ and $c_\alpha=  0.804$. }
\label{fig1_5}
\end{figure}

Furthermore, we point out that, starting from the data reported in Table  \ref{tab1}, by the reconstruction of $c_\alpha$ by the cubic spline interpolation $p(\alpha)$, we are able to find the numerical solutions for any values of $\alpha$ with $0<\alpha<1$. In Fig.(\ref{fig1_5}), we report the numerical solutions obtained for $\alpha = 0.33$ with $c_\alpha=2.895$ and $\alpha =0.66$ with $c_\alpha=0.804$.
Then, we can conclude that the latest numerical results confirm the validity of the analytical approach and the efficiency of the numerical method used.

\subsection{Test 2}

\begin{figure}[!h]
\centering
\includegraphics[width=.4\textwidth]{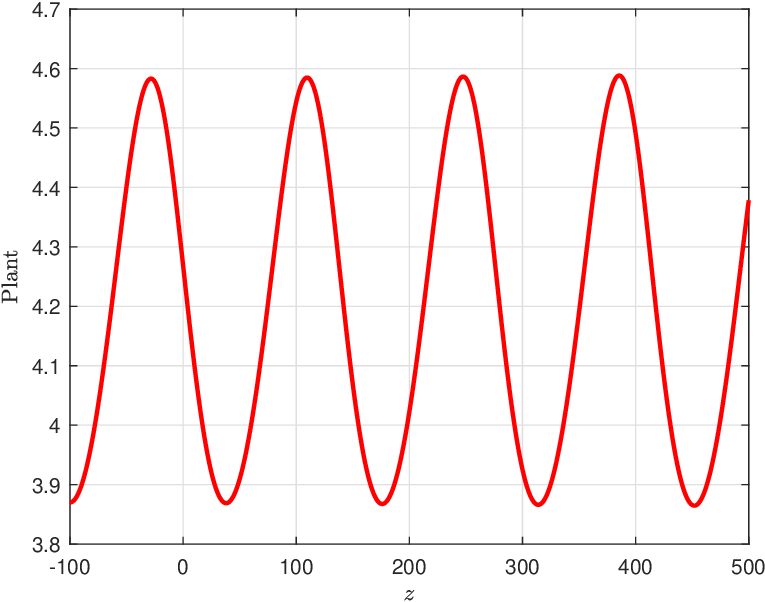} \quad
\includegraphics[width=.4\textwidth]{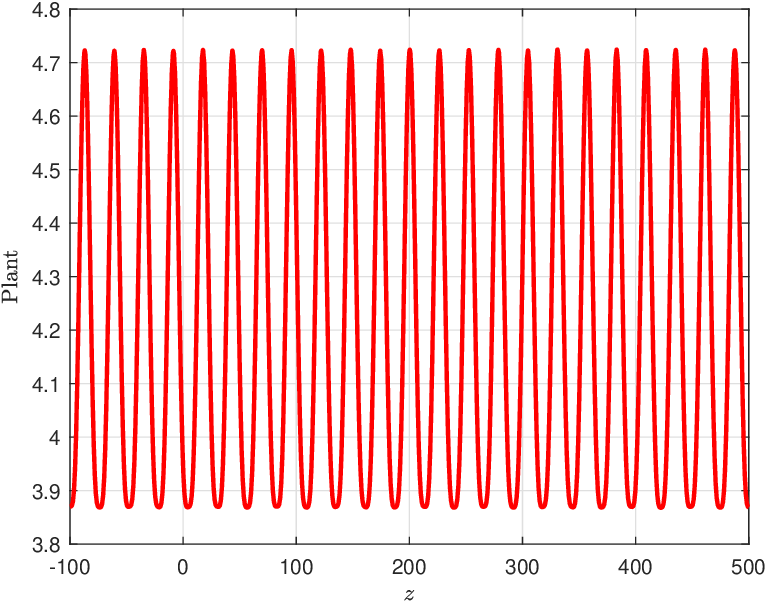} 
\caption{Test 2. Numerical solutions $U_j$ of the concentration of plant $U(z_j)$. Left frame: reduced KL model. Right frame: reduced KL--GS model.}
\label{fig2_0}
\end{figure}

\begin{figure}[!h]
\centering
\includegraphics[width=.45\textwidth]{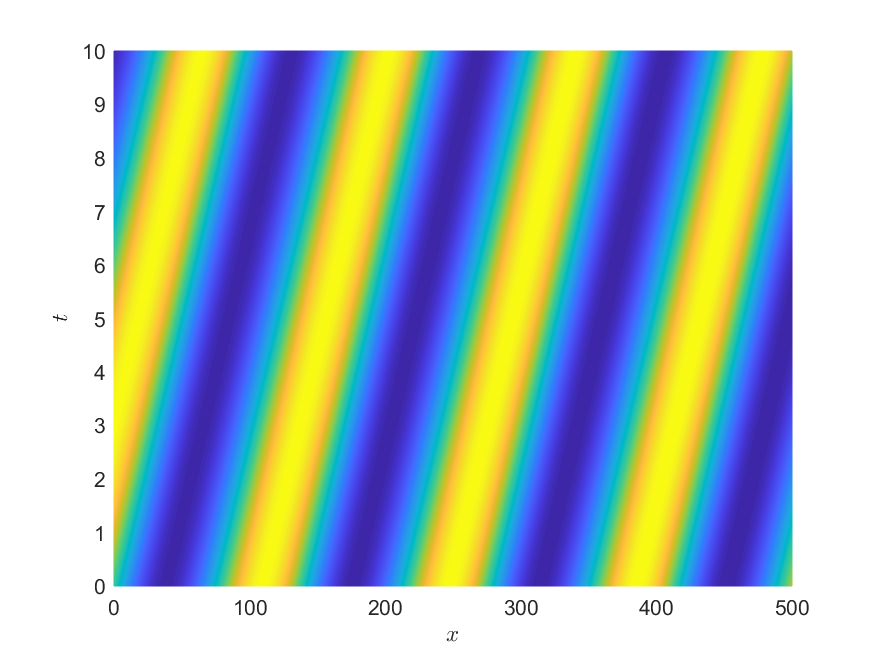}  \quad
\includegraphics[width=.45\textwidth]{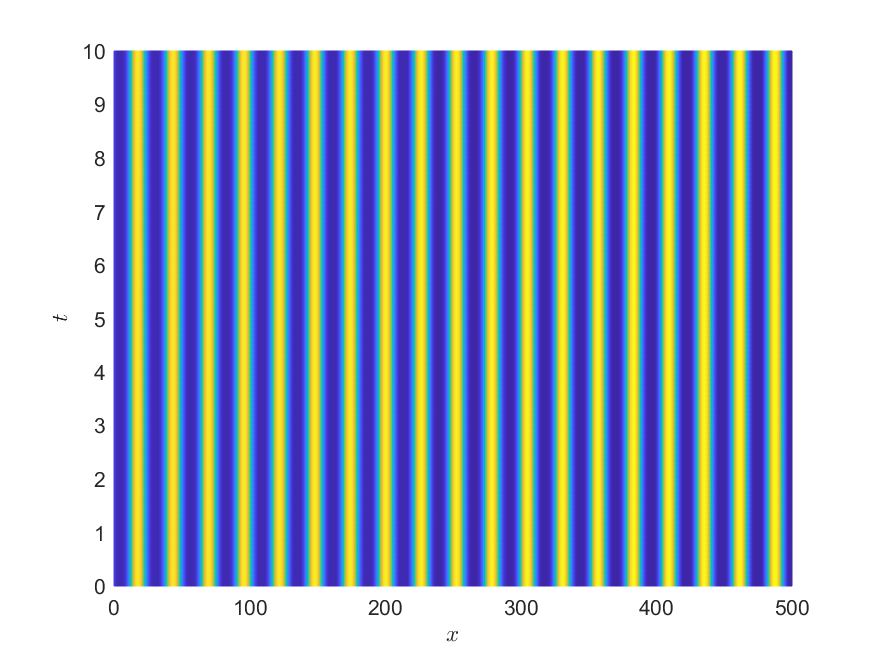} 
\caption{Test 2. Numerical solutions of the concentration of plant $u(t,x)$. Left frame: KL model. Right frame: KL--GS model.}
\label{fig2_00}
\end{figure}

In this second test, we change only the value of $\nu  =  380$. In Fig. \ref{fig2_0}, we report only the numerical approximation $U_j$ of $U(z_j)$ for $\alpha=0$, with $c = 9.313 <  9.37162 = c^{HB} $ and for $\alpha=1$ with $c=0$,  respectively. 
The vegetation patterning processes, on sloped and no--sloped terrains described by the KL and KL--GS models, are shown in  Figure \ref{fig2_00}. We report only the numerical solution  $u^n_j$, approximation of $u(t^n,x_j)$, reconstructed by using the cubic spline interpolation to evaluate them at any point of the computational domain, taking into account that $z=x- c t$ with final time $t=8$.

In the following, we choose $c_\alpha$ to obtain oscillatory solutions with a constant wavelength, as reported in Table \ref{tab2}.

\begin{figure}[!h]
\centering
\includegraphics[width=.45\textwidth]{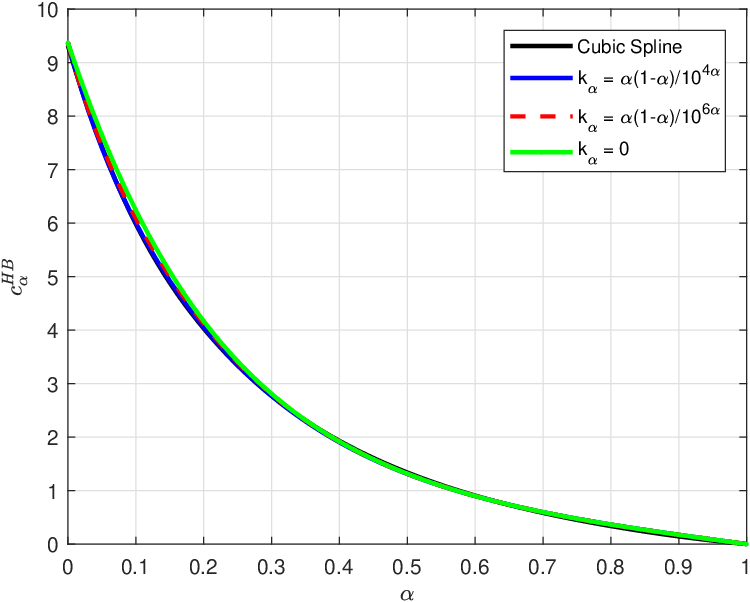} \qquad
\includegraphics[width=.45\textwidth]{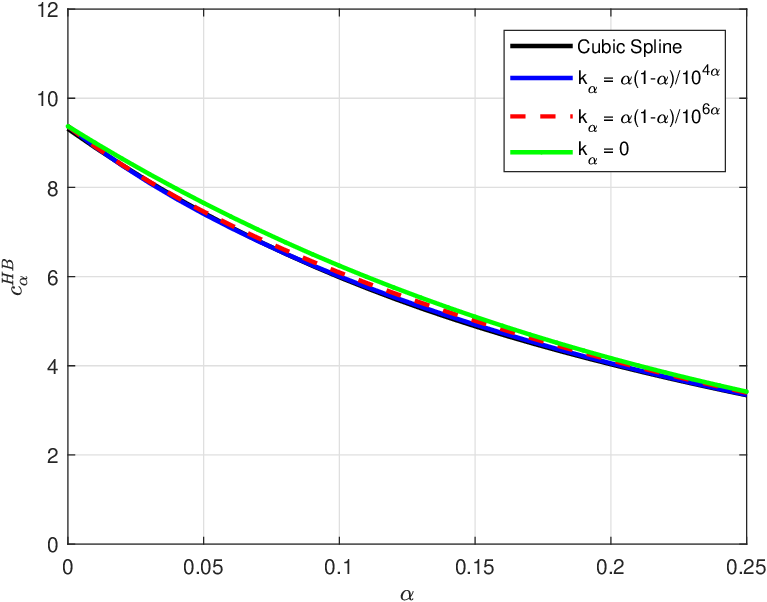} \caption{Migration speed $c^{HB}_\alpha$, depending on $\alpha$, given by (\ref{condHBF}) for $\nu=380$. Right frame: zoom of the left frame.}
\label{fig1_c380}
\end{figure}

\begin{table}[!h]
\begin{tabular}{|c|c|c|c|c|c|c|c|c|c|c|c|}
\hline
$\alpha$ &  0.0 & 0.1 & 0.2 & 0.3 & 0.4 & 0.5 & 0.6 & 0.7 & 0.8 & 0.9 & $\rightarrow$ 1 \\
\hline 
$c_\alpha$ & 9.313 & 5.985 & 4.032& 2.772& 1.929& 1.337& 0.906 & 0.583& 0.337 & 0.143  & $\rightarrow$ 0   \\
\hline 
\end{tabular}
\caption{Values of the migration speed $c_\alpha$ depending on fractional parameter $\alpha$.}  
\label{tab2}
\end{table}
In Fig. \ref{fig1_c380}, we show the migration speed $c_\alpha^{HB}$,  with $d=10^4$ (blue line), $d=10^6$ (red line) and $d \rightarrow \infty$ (green line), and the cubic spline interpolation (black line).

\begin{figure}[!h]
\centering
\begin{subfigure}
\centering
\includegraphics[width=.25\textwidth]{Test380_P_KL_2d.eps} \quad
\includegraphics[width=.25\textwidth]{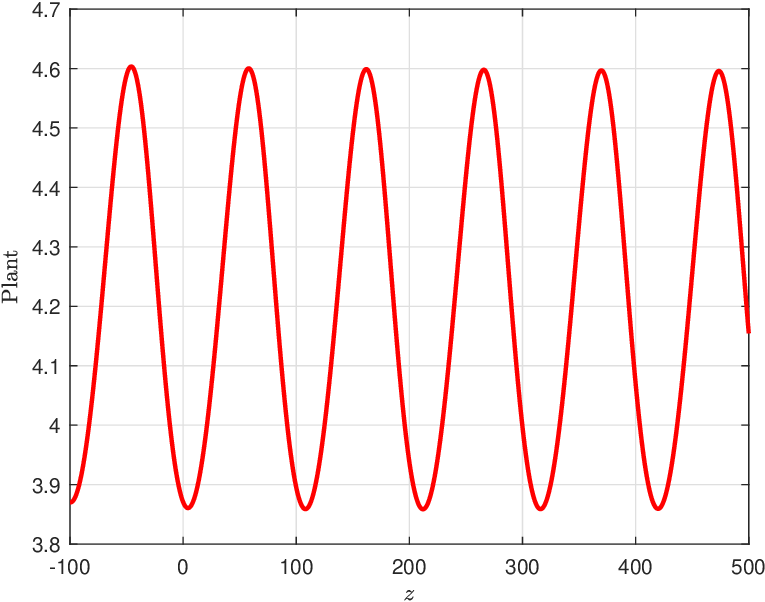} \quad
\includegraphics[width=.25\textwidth]{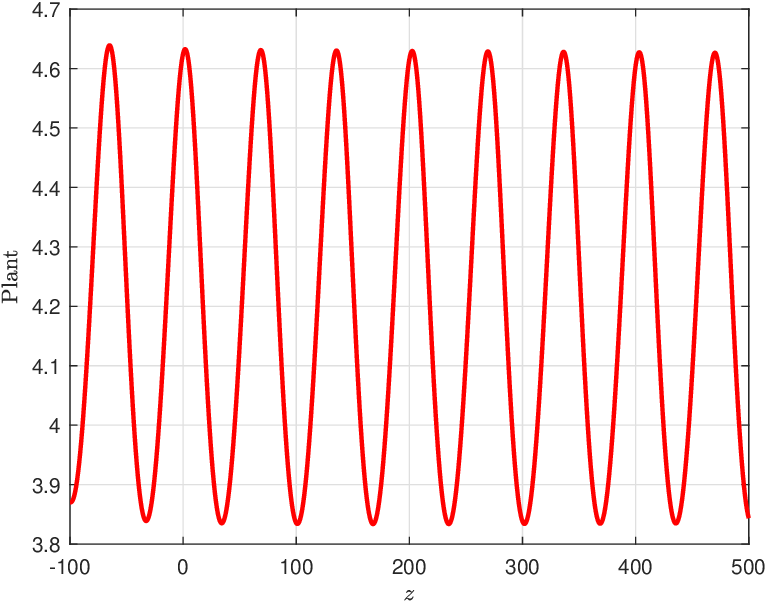} \\
\includegraphics[width=.25\textwidth]{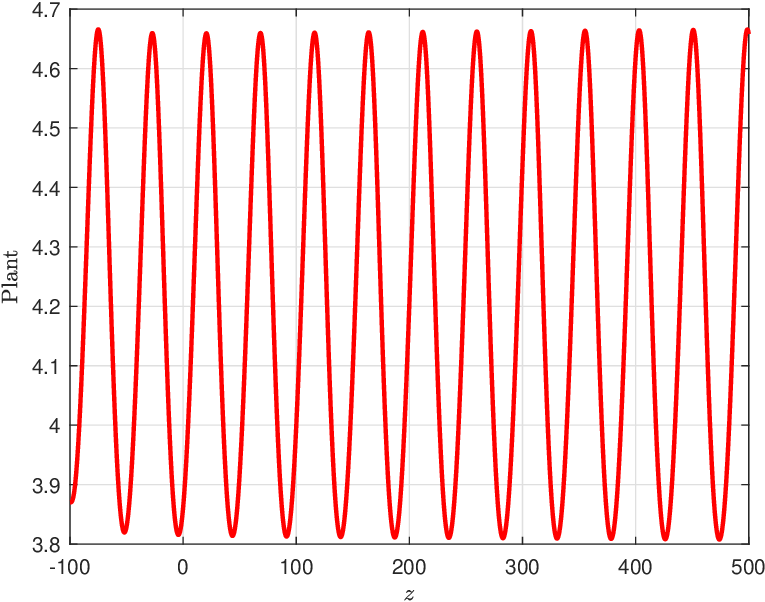} \quad
\includegraphics[width=.25\textwidth]{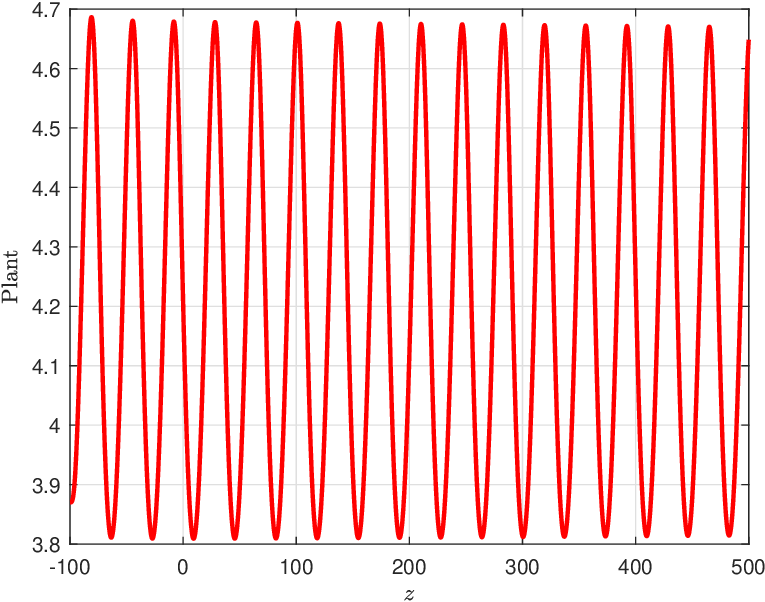} \quad
\includegraphics[width=.25\textwidth]{Test380_P_GS_2d.eps} 
\caption{Test 2. Numerical solutions for the concentration of plants, $U(z)$. From top to bottom: $\alpha= 10^{-8}$, $\alpha=0.1$,
$\alpha =0.3$, $\alpha=0.5$, $\alpha =0.7$ and $\alpha = .999999$.}
\label{fig2_1}
\end{subfigure}
\bigskip
\begin{subfigure}
\centering
\includegraphics[width=.28\textwidth]{Test380_P_KL.eps}  \quad
\includegraphics[width=.28\textwidth]{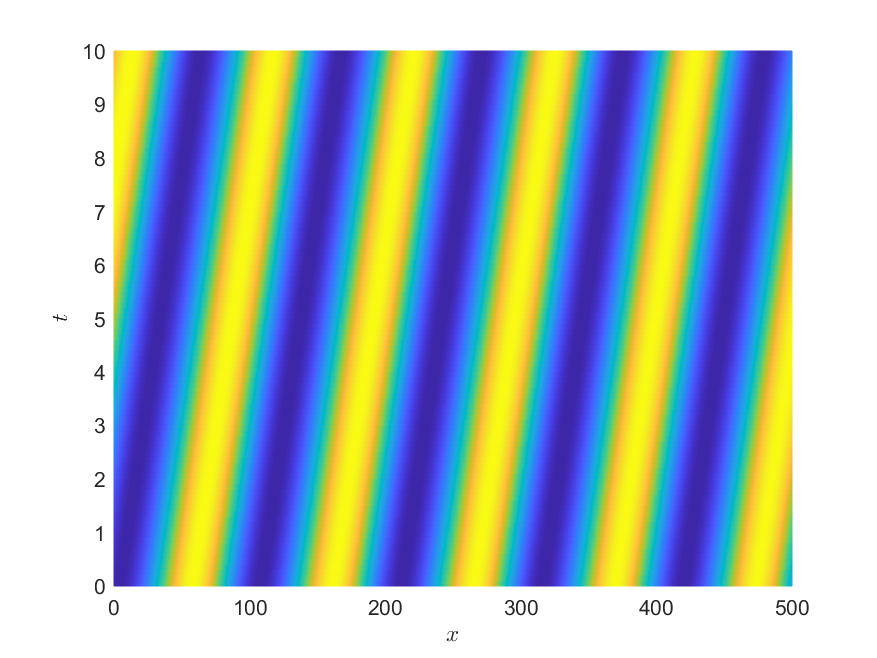} \quad
\includegraphics[width=.28\textwidth]{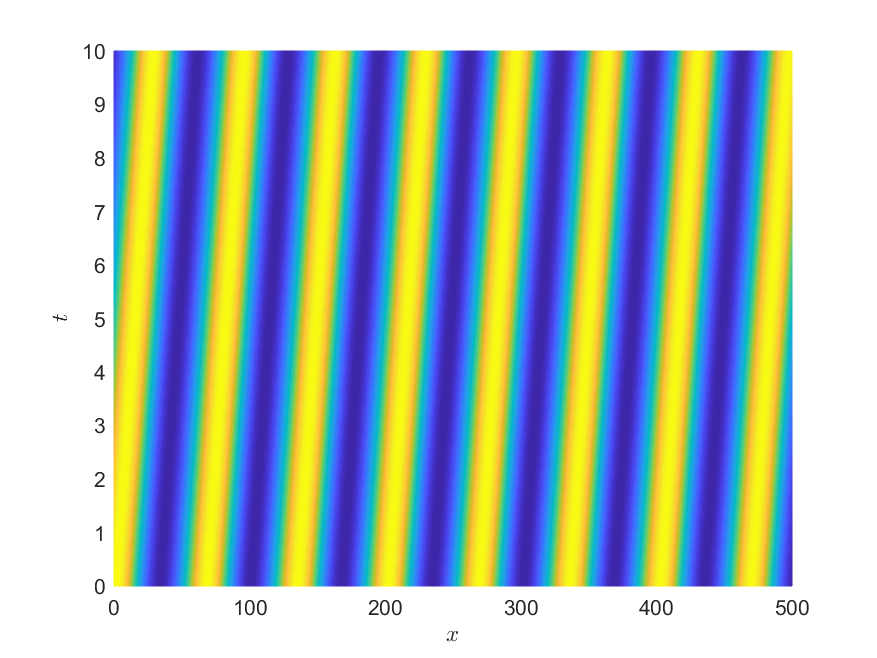} \\
\includegraphics[width=.28\textwidth]{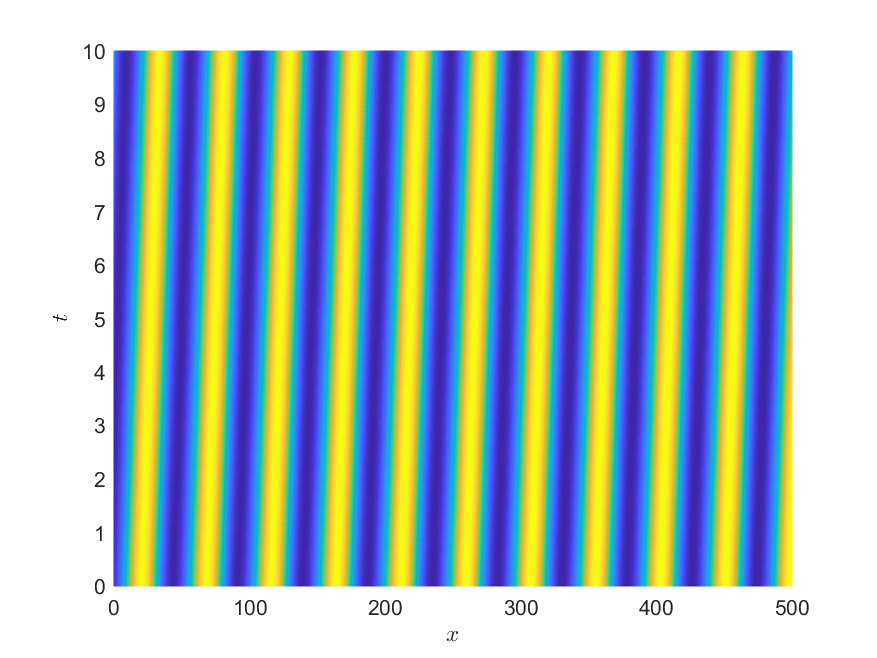} \quad
\includegraphics[width=.28\textwidth]{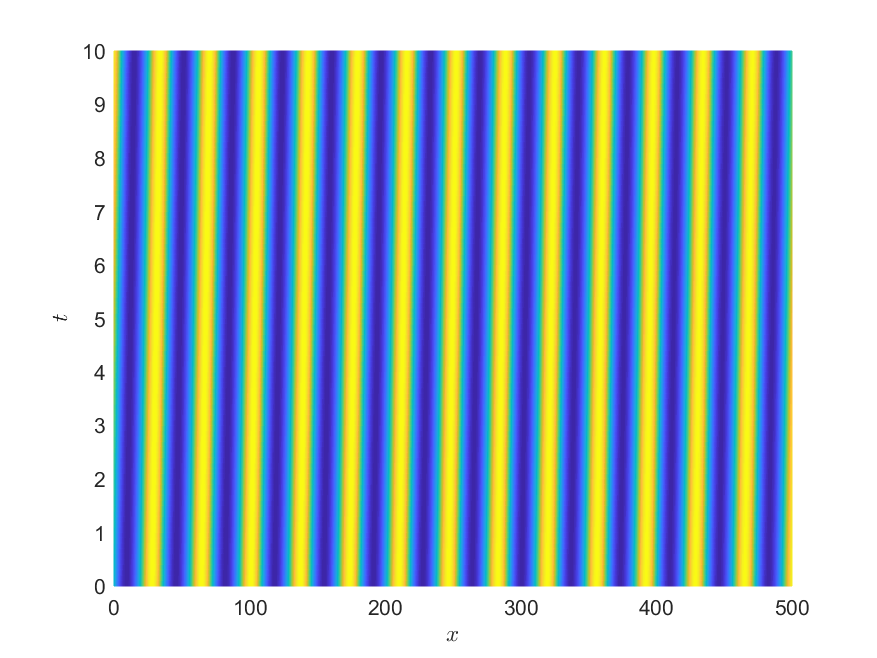} \quad
\includegraphics[width=.28\textwidth]{Test380_P_GS.eps} 
\caption{Test 2. Numerical solutions for the concentration of plants, $u(t,x)$, at final time $t=10$. From top to bottom: $\alpha= 10^{-8}$, $\alpha=0.1$, $\alpha =0.3$, $\alpha=0.5$, $\alpha =0.7$ and $\alpha = .999999$.}
\label{fig2_3}
\end{subfigure}
\end{figure}

\begin{figure}[!h]
\centering
\includegraphics[width=.34\textwidth]{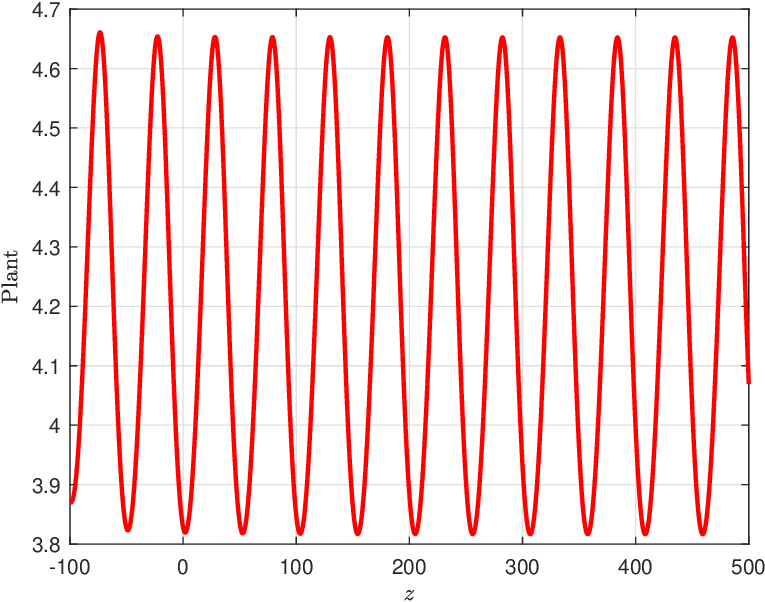} \qquad
\includegraphics[width=.38\textwidth]{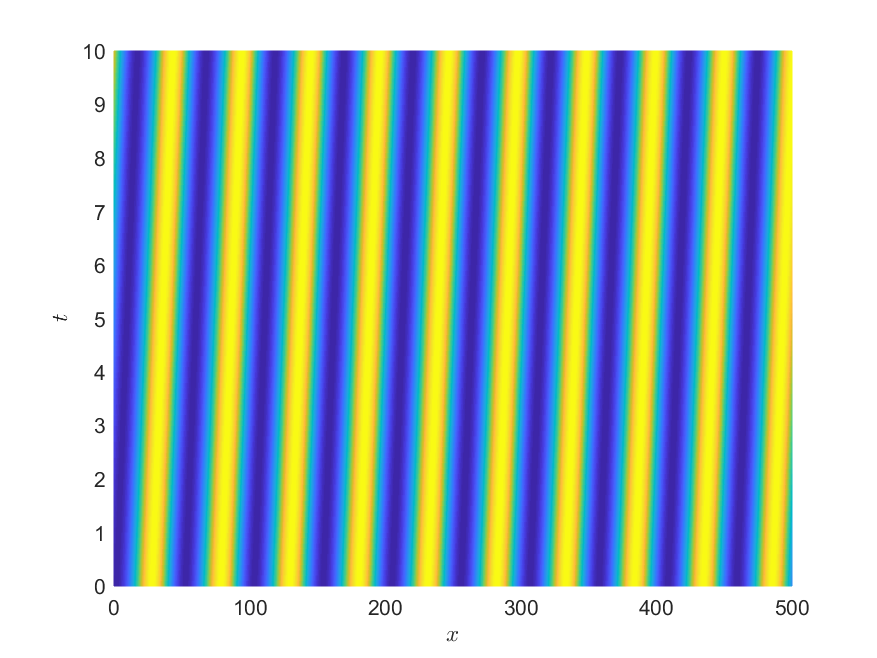} \\
\includegraphics[width=.34\textwidth]{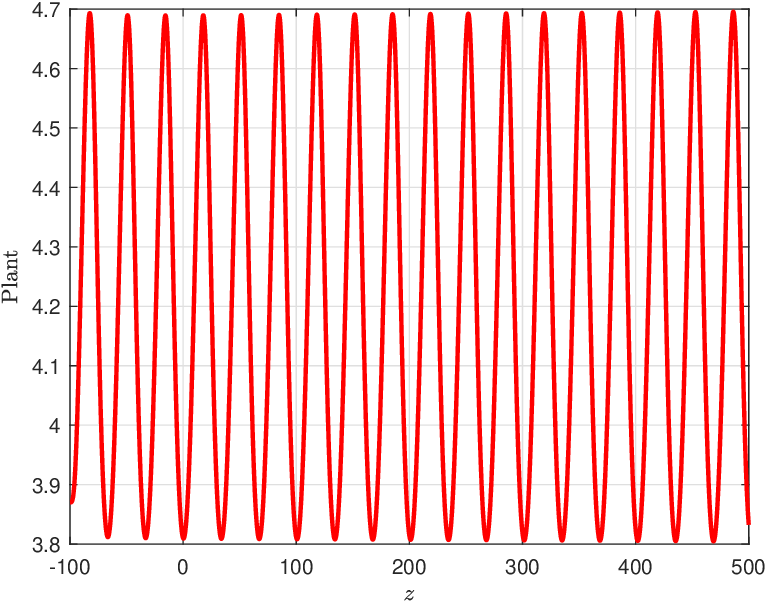} \qquad
\includegraphics[width=.38\textwidth]{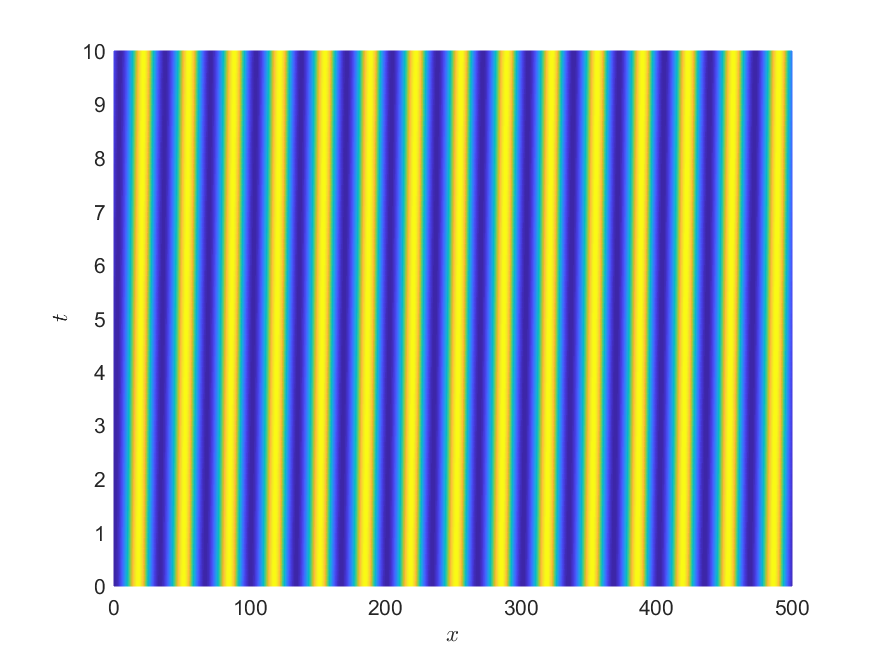} 
\caption{Test 2. Numerical solutions  $U^n_j$ of the plants $u(t,x)$ obtained for $a=2$, $m=0.45$, $ \nu = 380$. Top: $\alpha=0.46$ and $c_\alpha=  1.5504$. Bottom: $\alpha=0.77$ and $c_\alpha=  0.4046$. }
\label{fig2_4}
\end{figure}

In Figures \ref{fig2_1} and \ref{fig2_3}, we report only the numerical approximation $U_j$ and $u^n_j$ of $U(z_j)$ and $u(t^n,x_j)$, respectively, obtained for different values of $\alpha$.
Also for this example, the same conclusions arise: we note the 
differences in the vegetation patterns with respect to the slope of the domain and, then, with respect to different values of $\alpha$.
As in the previous example, starting from the cubic spline interpolation, in Fig. \ref{fig2_4}, we show the  oscillatory solutions, and the pattern for 
$\alpha=0.46$ with $c_\alpha=1.5504$ and $\alpha=0.77$ with $c_\alpha=0.4046$, respectively. These latest numerical results and those of Test 1 confirm the validity of the analytical approach and the efficiency of the numerical method used.

\bigskip

{\bf Remark 2. }  
In the KL model, on the no-flat domain, the migration of the plant biomass arises due to the advection term that describes the downward-oriented flow of the water.
 When $\alpha=1$, in the KL--GS model, the water does not flow, there is not advection term but the water diffusion process occurs. In this case, the solutions do not migrate but only diffuse.
 In the fractional model, as the $\alpha$ parameter increases, from zero to one, the trajectory of the solution changes due to the decrease of the migration speed related to the decreasing of the slope of the domain that tends to become flat, The differences in the trajectories of the vegetation patterns depend on the value of the migration speed $c_\alpha$, a function of the fractional parameter $\alpha$ that is linked to the slope of the domain.

\section{Conclusions}
In this article, we present a new fractional system to describe the interaction and dynamics between plants and water in non-flat and low-lying, arid and semi-arid environments.
The main goal of the proposed formulation lies in assuming that the parameter of the fractional operator is linked to the slope of the domain { so that the new fractional model can describe how the migration changes for different slopes and different values of $\alpha$. The new proposed fractional model is a connection between the KL model and the KL--GS one, as the parameter $\alpha$ varies.
This assumption is validated by the analytical study performed on the Hopf bifurcation of the migration speed, which shows how the migration speed $c$ depends on the fractional parameter $\alpha$.
The Hopf bifurcation point of the migration velocity is determined as a function of the parameter $\alpha$; so that the pattern bands leave from the Hopf bifurcation $c_\alpha^{HB}$. 
The fractional formulation of the mathematical model allows obtaining the oscillatory behavior of the solutions and, therefore, guarantees the formation of vegetation patterns in arid and semi-arid environments with any slope.} The numerical simulations confirm the self-organization of the plants into stripes whose migration depends on the speed $c_\alpha$, with $c_\alpha<c_\alpha^{HB}$, linked to the slope of the domain. The reported numerical results validate the theoretical results.

{ The analytical and numerical study of a new two--dimensional fractional model will represent the future direction of the research. Moreover, the numerical methods, implemented in \cite{Moniri_bis_2025} or in \cite{Desiderio_1,Desiderio_2}, could be used to compare the obtained solutions in this study.}
\bigskip

\section*{Acknowledgements}
A.J. and M.P.S are members of the GNCS-INdAM Research Group and of the GNFM-INdAM Research Group, respectively.

This research is supported by the project: "Strategie HPC e modelli fisico-matematici per la previsione di eventi meteorologici estremi (HPC-XTREME)". CUP 
B83C22002830001 - codice identificativo CN00000013. Bando PRIN 2022 PNRR - D.D. n. 341 del 15/03/2022.

\end{document}